\documentclass[12pt, letterpaper]{article}

\usepackage{amsmath,amssymb}
\usepackage{cite}
\usepackage{fancyhdr}
\usepackage[top=1in, bottom=1in, left=1in, right=1in]{geometry}
\usepackage[dvipdfm, dvips]{graphicx}
\usepackage{hyperref}
\usepackage{epsfig}
\usepackage{subfigure}

\numberwithin{equation}{section}

\begin{document}


\setcounter{page}{0}
\date{}

\lhead{}\chead{}\rhead{\footnotesize{RUNHETC-2011-22\\SCIPP-11-08\\UTTG-26-11/TCC-030-11}}\lfoot{}\cfoot{}\rfoot{}

\title{\textbf{Holographic Theories of Inflation and Fluctuations\vspace{0.4cm}}}

\author{Tom Banks$^{1,2}$\vspace{0.7cm}\\
{\normalsize{$^1$NHETC and Department of Physics and Astronomy, Rutgers University,}}\\
{\normalsize{Piscataway, NJ 08854-8019, USA}}\vspace{0.2cm}\\
{\normalsize{$^2$SCIPP and Department of Physics, University of California,}}\\
{\normalsize{Santa Cruz, CA 95064-1077, USA}}\vspace{0.7cm}\\
Willy Fischler$^{3}$\vspace{0.7cm}\\
{\normalsize{$^3$Department of Physics and Texas Cosmology Center, University of Texas,}}\\
{\normalsize{Austin, TX 78712, USA}}
}

\maketitle

\thispagestyle{fancy}

\begin{abstract}
\normalsize \noindent The theory of holographic space-time (HST)
generalizes both string theory and quantum field theory. It provides
a geometric rationale for supersymmetry (SUSY) and a formalism in
which super-Poincare invariance follows from Poincare invariance.
HST unifies particles and black holes, realizing both as excitations
of non-commutative geometrical variables on a holographic screen.
Compact extra dimensions are interpreted as finite dimensional
unitary representations of super-algebras, and have no moduli. Full
field theoretic Fock spaces, and continuous moduli are both emergent
phenomena of super-Poincare invariant limits in which the number of
holographic degrees of freedom goes to infinity. Finite radius de
Sitter (dS) spaces have no moduli, and break SUSY with a gravitino
mass scaling like $\Lambda^{1/4}$. We present a holographic theory of inflation
and fluctuations. The inflaton field is an emergent concept, describing the geometry
of an underlying HST model, rather than ``a field associated with a microscopic string theory".
We argue that the phrase in quotes is meaningless during the inflationary era, in the HST formalism.
\end{abstract}


\newpage
\tableofcontents
\vspace{1cm}


\section{Introduction}

The theory of Holographic Space-time is a generalization
of both quantum field theory (QFT) and string theory. It shares with
quantum field theory the assignment of an operator algebra to each
causal diamond in space-time, and with string theory the holographic
nature of its degrees of freedom and its dependence on supersymmetry
(SUSY).

Although it is not usually stated this way, QFT already encodes the
causal structure of space-time into the structure of the quantum
operator algebra. This is somewhat obscured by the fact that the
operator algebras in QFT are all infinite dimensional and often have many
inequivalent unitary representations. The detailed algebraic
structure depends on which QFT is under discussion. By contrast, if
one adopts the strong form of the holographic principle advocated by
the authors\cite{holocosm}\cite{bousso}, then the algebras of finite area
causal diamonds\footnote{Area here refers to the area of the {\it
holographic screen} of the diamond: the maximal area space-like
$d-2$ surface on its null boundary.} are finite dimensional.
Furthermore, this can be viewed as a way of encoding the conformal
factor of the geometry in the structure of the quantum operator
algebra. As a consequence one can state the properties of the
holographic theory entirely in terms of quantum mechanics, and the
space-time interpretation of the formalism is {\it emergent} in the
limit of large causal diamonds.

The first step in constructing a quantum space-time is thus a
collection of matrix algebras $A(D_i)$\footnote{Each of the algebras
$A(D)$ is a factor.}, where the label on the index set reminds us
that these will eventually be associated with causal diamonds. For
each pair of algebras we identify a common tensor factor $O_{ij} =
O(D_i , D_j )$, which is the algebra associated with the largest
causal diamond in the intersection of the two.
\subsection{Quantum mechanics and observers}

We put a little more structure on the collection of algebras by introducing
Hilbert spaces of time-like observers, or simply of time-like trajectories. An observer in quantum mechanics
is a quantum system with a large collection of semi-classical observables. The notion of
semi-classical depends on two criteria, both of which are only satisfied by systems with a large
number of states. The first is that the dynamics of the system must contain a small parameter,
which guarantees that the WKB approximation is valid for the observables in question. This parameter
is typically proportional to the inverse logarithm of the number of states of the system, which must therefore
be exponentially large. Furthermore, each semi-classical state of the observable in question must correspond
to an exponentially large number of micro-states, and the dynamics of the system must be such that rapid transitions among these micro-states occur on a time scale fast compared to the classical motion of the observable. This criterion ensures rapid decoherence of quantum phases correlating different semi-classical states of the observable.

In models of quantum gravity, there are two kinds of subsystems, which have such semi-classical observables. The first consists of collections of particles, whose interactions are well modeled by a cutoff effective quantum field theory. These subsystems have many semi-classical observables, given by field averages over large subsets of the volume described by the field theory, if that volume is large in cutoff units. These provide pointers and dials on ''measuring apparatus", which can be constructed from particle degrees of freedom. The second important class of subsystems consists of black holes. No Hair theorems tell us that, while these systems have many quantum states, they have very few semi-classical observables. Either kind of ``observer" will follow a time-like trajectory through space-time.

Geometrically, we can think of the time-like trajectory of an observer as being equivalent to a nested sequence of causal diamonds, corresponding
to the causal diamonds defined by pairs of points $P(n), Q(n)$ with larger and larger time-like separation along the trajectory.
This corresponds to a nested sequence of Hilbert
spaces ${\cal H} (n , x) = {\cal P}^{N(n)} $ \footnote{As we will discuss in a minute, $x$ labels different
time-like trajectories in a congruence}. For reasons that will
become apparent below, we call ${\cal P}$ the {\it single pixel
Hilbert space}. If space-time has a beginning, as we expect
for a Big Bang universe, it is convenient to think of causal
diamonds whose past tip lies on the Big Bang hypersurface. If it
has time reversal symmetry, we might prefer to take segments that
penetrate a space-like surface invariant under T inversion, and
extend in a time symmetric way to its past and future. In the latter
case, dynamics will be described by a sequence of {\it partial
S-matrices}, unitary maps $U(T,-T)$, from the past boundary of each causal
diamond to its future boundary. For a cosmological space-time we
have ordinary time dependent Hamiltonian evolution, starting at the
Big Bang.

If we think of the time differences in this sequence as the Planck time, then we expect that for large integer $n$ , $N(n) \sim n^{(d - 2)}$ in $d$ space-time dimensions. This is the way the area of a causal diamond
scales. For $d=4$ we will see that this scaling is natural, and resolves a discrepancy between quantum and geometrical descriptions of the $p=\rho$ FRW space-time, which was uncovered in \cite{holocosmbfm}.

To incorporate the concept of particle horizon, or causality more generally, on the dynamics, we
insist that the unitary transformation that takes one from the Big
Bang to some value of $n$, is a tensor product of a unitary in
${\cal H} (n,x) $ and a unitary in its tensor complement in the
larger space ${\cal H} (n_{max} (x), x)$. $N(n_{max}) (x) {\rm ln\
dim\ }{\cal P}$ is $4$ times the area of maximal area causal diamond
encountered by the observer labeled by $x$\footnote{This statement
is meant in an asymptotic sense. There are corrections to the entropy area relation.}. Notice that the
Hamiltonian {\it must} be time dependent in order for this
construction to make sense.
Now we must specify the overlaps ${\cal O} (n; x, y)$ for every pair
of Hilbert spaces. We do this by giving the space of $x$ labels the
topology of some, not necessarily regular, lattice in infinite flat
$d-1$ dimensional space. For nearest neighbors on the lattice we
insist that the overlap be ${\cal P}^{N(n-1)}$. For other pairs of
points we insist that the dimension of the overlap not increase as
$d(x,y)$ increases. $d(x,y)$ is the minimum number of lattice steps
between $x$ and $y$. The rest of the overlaps are specified as part
of the solution to a basic dynamical consistency condition, which
follows from the formalism. {\it The Hamiltonian evolutions of two
different time-like observers, together with specified initial data,
each determine a density matrix in every overlap space. These
density matrices must be related by a unitary conjugation, for every pair of points, and for every time}. This
is an infinite number of complicated conditions, reminiscent of the
concept of {\it many fingered time}, and only a few solutions to
them are known\footnote{The connection to many fingered time: Traditional Hamiltonian gravity introduces a local time evolution operator $H(x)$ and a consistency condition that the Poisson bracket of $H(x)$ with itself at two different space points is just a spatial diffeomorphism and therefore physically trivial. Here we introduce Hamiltonians for each time-like observer, and a consistency condition that on regions of space-time that are described by two different observers, the (impure) states predicted by the different Hamiltonian evolutions are quantum mechanically equivalent.} . It is not unreasonable to guess that these conditions are enough to determine completely the notion of a
quantum space-time, but perhaps other axioms will have to be added. Note that while the structures of the operator algebras in QFT and HST are similar, the Hamiltonian quantum mechanics is completely different. QFT has a Hamiltonian which is a sum of terms coming from different points of space. In HST Hamiltonians are associated to time-like trajectories. The Hamiltonian of any given trajectory is a complete description of the universe and describes all possible observations made by that observer. The fundamental dynamical principle is a consistency condition between the descriptions of the universe used by different observers. This is a generalization of the idea of Black Hole Complementarity\cite{sussuglum}.

A number of observations are in order:
\begin{itemize}

\item The formalism obviously involves a choice of coordinates. From
the geometrical point of view, we have chosen a Cauchy surface and a
set of non-intersecting time-like trajectories intersecting that
surface, to define a coordinate system on space-time. This was
inevitable, since we are constructing a unitary quantum theory of a
gauge theory. The holographic principle sheds new light on the
principle of general covariance. A system localized inside a finite
area causal diamond has only a finite number of states. Observers in
QM are large systems with many independent observables that behave
semi-classically. Thus, there can be no quantum mechanically ideal
local observers. This means that there is an unavoidable ambiguity
in the theoretical description of a local region. Two different
choices of Hamiltonian, which give results that are equivalent
within the intrinsic limits of accuracy of local measurement, have
no measurable distinction. Only idealized measurements on the
boundaries of infinite space-times can be defined with unambiguous
mathematical precision. This is exactly the only kind of gauge
invariant data provided by string theory in asymptotically flat or
AdS space.

\item The time slicing used in our coordinate system is ``equal
area" slicing: simultaneous observers are those whose holographic
screens have the same area. We will introduce a somewhat different slicing in our
attempt to construct a realistic cosmology below.

\item {\it Space-time geometry is not a fluctuating quantum
variable.} Rather it is encoded in the overlap rules, which are
themselves dependent on, and constrain, the Hamiltonian. A widely
ignored hint that this might be the case comes from the
Wheeler-Dewitt approach to quantum gravity (which, in our opinion
makes sense {\it only} in the semi-classical expansion). Each
solution of the WD equation determines a different time dependent
quantum Hamiltonian, correlated with a different classical
space-time.

\end{itemize}

The last item suggests a question, ``What are the variables of the
quantum theory of gravity?'', to whose answer we turn in the next
section.

We want to emphasize that our construction can be phrased entirely in terms of quantum mechanics. It contains time as a fundamental concept, but space, and Lorentzian space-time are emergent. It is an infinite collection of quantum systems, each obeying the causality requirement we would expect for the Hamiltonian of a system on a time-like trajectory: its time dependent Hamiltonian couples together only a subset of the degrees of freedom at
any time, and that subset grows with time, perhaps reaching an infinite dimensional Hilbert space. The growth is by addition of tensor factors so that ``older" operators commute with newer ones. The different systems are related by a set of overlap rules, which imply that the dynamics imposed by any pair of them on the tensor factors they share in common throughout history, is the same up to a change of basis. The index set labeling the different systems has the topology, but not the geometry of a lattice. We can think of this lattice as the definition of the topology of a space-like Cauchy surface for the emergent space-time. The space-time geometry is encoded in the overlap rules, which supply both the conformal factor and the causal structure of the emergent metric, for large Hilbert spaces, which correspond to large causal diamonds. The Bekenstein-Hawking area law is built into our construction, so that, following Jacobson\cite{jacobson}, we can assert that the geometry satisfies Einstein's equations, with a stress tensor whose integrals are related to the thermodynamic averages of the Hamiltonian of local Rindler observers, with infinite acceleration\footnote{In actual fact, the temperature is infinite, but the local Hilbert spaces are finite dimensional, so one should think of a maximally accelerated observer on a stretched horizon.}.

\section{SUSY and the holographic screens}

The areas that we encode as dimensions of Hilbert spaces, are areas
of holographic screens. Consider the orientation of an infinitesimal pixel on such
a screen, in d dimensional space-time. It is determined by a null
vector, and a plane transverse to it. This information can be
encoded in a solution of the Cartan-Penrose (C-P) equation
$$\bar{\psi} \gamma^{\mu} \psi (\gamma_{\mu} )^{\alpha}_{\beta}
\psi^{\beta} = 0.$$ It follows from this equation that the spinor
bilinear $n^{\mu} = \bar{\psi} \gamma^{\mu} \psi$ is null, and that
the spinor itself is a null plane spinor for that null vector. That
is, the bilinears $\bar{\psi} \gamma^{\mu_1 \ldots \mu_k} \psi$, for
$k \geq 2$, all lie in hyper-planes embedded in the same space-like
$d-2$ plane, transverse to $n^{\mu}$.

These variables thus lie in the spinor bundle over the holographic
screen. We choose a gauge for the local Lorentz invariance of the
C-P equation so that the direction of the null vector is determined
by the coordinate $\xi$ on the screen and the spinors are $S_a (\xi
)$, where $S_a$ is a $d-2$ spinor. The simplest example
is a spherical holographic screen, where the null direction is $(1, {\bf \Omega})$, with ${\bf \Omega}$
a unit $d-1$ vector. Note that there are two inequivalent choices of the orientation of ${\bf \Omega}$,
which, for asymptotically flat space gives us incoming and outgoing states.

The holographic principle tells
us that these operators must be quantized in such a way that their
representation space is a finite dimensional Hilbert space. The
first implication of this is that the space of sections of the
spinor bundle must be finite dimensional, which means that the
holoscreen cannot be thought of as a normal manifold. We choose
instead to think of it as a fuzzy manifold. The topology of any
manifold is completely encoded in the C* algebra of its continuous
functions. One can approximate this algebra (or the algebra of
smooth or measurable functions) by viewing it as a limit of algebras
of matrices. Geometric quantization of compact symplectic manifolds
provides a wealth of examples. We will illustrate the technique for
a simple, but very relevant, example.

$SU(2)$ has a $K$ dimensional representation for every $K$. It
follows that the space of $K \times K$ matrices is a reducible
representation of $SU(2)$, containing all spins up to $K-1$. If we
define $$\Omega^a \equiv \frac{J^a}{\sqrt{\frac{K^2 - 1}{4}}},$$ then
in the limit $K \rightarrow\infty$ these become commuting
coordinates on the two-sphere. We get, measurable, continuous, or
smooth functions on the sphere by considering matrices whose
expansion in terms of $SU(2)$ representations has the appropriate
falloff for large spin. The two chiral spinor bundles over the
fuzzy two sphere are a complex $K \times K + 1$ matrix $\psi_i^A$
and its adjoint $\psi^{\dagger\ j}_B $. The unique $SU(2)$
covariant quantization rule for these variables, with a finite
number of states, is

$$[\psi_i^A , \psi^{\dagger\ j}_B ]_+ = \delta_i^j \delta^A_B .$$
We can take \cite{bfm} a limit as $K \rightarrow \infty$ such that
these operators approach an operator valued measure $\psi (\Omega,
\Omega_0 ) $ on the space of measurable sections of the chiral
spinor bundle on the two sphere

$$\Psi [q] = \int\ d\Omega \psi (\Omega , \Omega_0 ) q(\Omega ) = q
(\Omega_0 ) p \psi , $$ where $[\psi, \psi^{\dagger}]_+ = 1$, and $p$ is a positive real normalization constant that appears in taking the $K\rightarrow\infty$ limit.The commutation relations imply
$$[\psi [q], \psi^{\dagger} [q*] ]_+ = q* q (\Omega_0 ).$$ The spin cover of
the Lorentz group, $SL(2) \times SL(2)$, is the spin cover of the conformal
group of the two sphere. The conformal Killing spinor (CKS) equation
$$ D_A q^{\alpha} = \sigma_A q^{\alpha} ,$$ where $\sigma_A$ is the
2D Weyl vector dotted into the zweibein on the sphere, has two
solutions, which transform as a 4D Weyl spinor under $SL(2)\times SL(2)$. It
is easy to verify\cite{bfm} that $$Q^{\alpha} = \psi [q^{\alpha}]
,$$ and its complex conjugate, satisfy the SUSY algebra
$$[Q_{\alpha}, \bar{Q}_{\dot{\beta}} ]_+ =
\sigma^{\mu}_{\alpha\dot{\beta}} P_{\mu} ,$$ where $$P_{\mu} = p (1,
\Omega_0 ),$$ with $p$ a positive real number that appears when
writing the $K \rightarrow\infty$ limit of the discrete commutation
relations in terms of measures. We can generalize this by choosing
our algebra of functions to be the direct sum of $L$ $K_i \times
K_i$ matrix algebras with its usual $S_L$ permutation gauge
invariance. We get $L$ massless supersymmetric particles with the
correct connection between spin and statistics.

It is natural to ask about the other matrices, which fill out the
algebra to that of $N \times N$ matrices, with $N = \sum K_i$. Let
us fix $N$ and try to maximize the entropy in states that could be
well described by a Fock space, as in quantum field theory. For
that, we need the individual $K_i$ to be large, so that each
``particle" can be well localized on the holographic screen, but we
also want to have many particles. The compromise is reached with
$K_i \sim \sqrt{N}$ and gives a total entropy $\sim N^{3/2}$,
whereas using the spinor bundle over the full $N \times N$ matrix
algebra would give an entropy $\sim N^2$. These scaling laws are
precisely those expected for the entropy of a field theory coupled
to gravity in a regions of size $R\sim N M_P^{-1}$, taking into
account the constraint that the back-reaction of the field theoretic
state does not produce black holes of size $\sim R$\cite{ckn}. If we
assume $K_i = 1$ corresponds to momentum $\sim 1/ R$, then we
reproduce the UV cutoff on field theory that is required to enforce
the back reaction constraint. Note however that in the holographic
formalism there is no such UV constraint on individual particles.
The momentum of each particle is constrained by all of the others. We can have a few particles
of high momentum, or many with low momentum. The lowest allowed momentum does
not really describe a particle, because it has no angular localization. The highest allowed momentum is of
order the Planck scale. If we try to enclose a particle of larger energy in a finite region, we will produce a black
hole instead.

The mathematics of the $N\rightarrow\infty$ limit involves the
unique Type $II_{\infty}$ von Neumann factor whose outer
automorphism group is the positive real numbers. These automorphisms
are essential to implementing boost invariance in the
limit\cite{11D}.

All of this talk of black holes assumes that we have constructed a
theory of quantum gravity. However, the commutation relations we
have postulated give rise instead to a massless chiral multiplet in
the $N\rightarrow\infty$ limit. The SUSY commutation relations for a
massless particle allow an arbitrary helicity for the ground state,
but our formalism preserves manifest $SO(3)$ invariance and does not
have that freedom. The maximum spin is tied to the spin of the
supercharges. The obvious solution to this is to increase the
number of pixel variables, and postulate a super-algebra

$$[(\psi^M)_i^A , (\psi^{\dagger\ P})_B^j ]_+ = \delta_i^j \delta^A_B Z^{MP}.$$
$M$ and $P$ label elements in a basis for sections of the spinor
bundle over a fuzzy compact manifold, while $Z^{MP}$ labels a basis
for the space of sections of the bundle of differential forms. We
will call the smallest irreducible unitary representation of this
algebra the {\it pixel Hilbert space}, ${\cal P}$. We will use the same symbol for the dimension of this Hilbert space.

Fuzzy compactification is well known for compact symplectic
manifolds, and even for odd dimensional manifolds with a Poisson
structure. While most of the manifolds so far encountered in string
theory are of this type, there is a more appropriate
formalism based directly on the spinor bundle\cite{tbjk}. We make the geometry fuzzy by restricting the space of solutions of the Dirac equation by a sharp eigenvalue
cutoff. The algebra of matrices in this finite dimensional spinor bundle, ``converges" to the algebra
of operators on the space of square integrable spinor sections, which contains the algebra of differential forms with Clifford multiplication as a subalgebra, as the eigenvalue cutoff is removed. As usual, convergence is achieved
by imposing restrictions on the form of the matrices in the infinite cutoff limit. Different restrictions produce the algebras of measurable, continuous, or smooth forms.

The {\it pixel superalgebra} should have a finite dimensional
unitary representation, with dimension $e^{\cal L}$. ${\cal L}$ is roughly the number of fermionic generators in the super-algebra. In the
$N\rightarrow\infty$ limit this representation should approach a
Fock space of massless super-particles including at least the ${\cal
N} = 1$ SUGRA multiplet, in the manner outlined for the chiral
multiplet above. The catalog of all algebras with this property is
the analog of the list of ``super-Poincare invariant
compactifications" in ordinary string theory. Note that in the spinor bundle
approach to fuzzy compactification, the notion of covariantly constant spinors is incorporated exactly in the fuzzy geometry. They give rise to zero modes of the Dirac equation. It's easy to verify that the pixel superalgebra automatically has a subalgebra, which converges to the super-Poincare algebra when $N \rightarrow\infty$, whenever such a covariantly constant spinor exists.

Using the principle that entropy is always one quarter of the area
of a $d-2$ surface in Planck units, we find
$$N^2 {\cal L} = \pi (RM_P)^2 = \pi R^2 V_D (M_D)^{D+2} ,$$ where
$M_D$ is the $D+4$ dimensional Planck mass. It's natural to
decompose this equation by identifying
$${\cal L} = (M_P / M_D)^2 = M_D^D V_D ,$$ since increasing ${\cal
L}$ increases the accuracy with which the fuzzy internal geometry
approximates a continuous one, while increasing $N$ plays the same
role for the two sphere. In fact, since the large eigenvalue spectrum of
the Dirac operator, with eigenvalue $\pm p$, always behaves as $p^D$, and this
is the number of independent fermionic generators of the pixel superalgebra,
${\cal L}$ will always scale like a $D$ dimensional volume.

Fixing ${\cal L}$ with $N\rightarrow\infty$ gives us a
super-Poincare invariant kinematics, with internal dimensions fixed
in Planck units. One of the remarkable features of this proposal
for compactification is that {\it it has no moduli}. The unitary
finite dimensional representations of a super-algebra are
characterized by discrete parameters. Compactifications with
continuous moduli can be achieved by taking ${\cal L}$ to infinity
with $N$\cite{tbjk}. In general there are many ways to do this and fixed
ratios of discrete parameters that go to infinity become continuous
moduli. From this point of view, moduli are an artifact of having
internal dimensions that are large in Planck units. One should not
think of them as intrinsically continuous variables, which must be
fixed by some dynamically generated potential. The implications of
this point of view for string inspired phenomenology will not escape
most of our readers.

Much of string cosmology, and phenomenological issues like the stringy origin of QCD axions depends on the idea that the massless scalars that appear in many string compactifications to Minkowski spaces of dimension less than eleven play important roles in the real world. In weakly coupled
string theory one can prove that most of these particles remain massless to all orders of perturbation theory. One then imagines an ``effective potential on moduli space", which will freeze the constant mode of the field but give rise to relatively light scalar particles, whose effective fields could play a role in inflation, the solution of the strong CP problem etc.

If holographic space-time is indeed the general theory to which existing string perturbation expansions are approximations, these ideas are wrong, or at least non-generic. Compact dimensions can exhibit continuous moduli in HST only when they are asymptotically large in Planck units. This is particularly important for cosmology. In the HST approach to Big Bang cosmology, early cosmological evolution takes place in a Hilbert space of relatively small dimension. There is no place for continuous moduli. Indeed, as we have seen, particle physics in HST arises, as in Matrix Theory\cite{bfss} in terms of block diagonal $N \times N$ matrices, where $N$ parameterizes the fuzziness of the two sphere.
In order to have a good field theoretic approximation, with both many particles and good localization for each particle, each block must be large and there must be many blocks. The CEB bounds the total entropy, by the area of the holographic screen of the particle horizon. This entropy is $N^2 {\rm ln} {\cal P}$, where ${\cal P}$ is the dimension of the smallest unitary representation of the pixel superalgebra. The number of fermionic generators of this algebra is of order the volume of the compactified space in Planck units, when that number is large. {\it Thus, for fixed entropy, if we insist on a field theoretic approximation to the dynamics of particles in the non-compact dimensions of space, we cannot also be in a regime where the compact space has approximate moduli.} When the cosmological evolution asymptotes to dS space, so that the total entropy is finite, this continues to be true throughout cosmological history.

We want to emphasize a couple of points that are implicit in the above discussion. We've talked about a general causal diamond, but also taken limits in which the conformal group of the two sphere is an emergent symmetry. Whenever the internal manifold has a covariantly constant spinor, this leads to super-Poincare invariant kinematics. In particular, if there is a graviton, then there must be a gravitino, and familiar low energy theorems tell us that the scattering matrix must be supersymmetric as well. Furthermore, the Minkowski space momentum operator is, if it exists at all, constructed out of a product of two fermionic generators of the pixel algebra. If there is no covariantly constant spinor\footnote{In \cite{tbjk} a flux generalization of the Dirac operator is proposed, which can describe supersymmetic flux compactifications, with spinors covariantly constant w.r.t. a generalized connection including coupling to the fluxes. } it is not clear how to obtain a momentum operator, even at the single particle level, let alone guarantee that momentum be conserved. While this is not yet a proof, it is a strong indication that the HST formalism will produce Poincare invariance only in combination with exact SUSY. We have empirical evidence from extant controlled constructions in string theory that this association between SUSY and Poincare invariance is in fact correct.

Conversely, if we are in dS space, which is described by a finite dimensional Hilbert space, we expect both SUSY and Poincare invariance to be broken. The dS radius is given in terms of $N$ and the dimension of the pixel Hilbert space by

$$\pi (RM_P)^2 = N^2 {\rm ln}{\cal P} \sim N^2 (M_P / M_D)^2 .$$ In the second part of this formula we are using Kaluza-Klein ideas, and $M_D$ is the higher dimensional Planck scale. We will take this to be the unification scale, $M_U \sim 2 \times 10^{16}$ GeV, following ideas of Witten\cite{wittenstrongcy}. The $\sim$ symbol reflects both geometrical factors in the relation between the GUT symmetry breaking scale and the actual higher dimension Planck scale, and, possibly, corrections to the KK picture of a smooth internal geometry. Given the nominal value of the c.c. in our world, this formula gives $N \sim \sqrt{\pi} (M_U /M_P ) 10^{61} \sim 4 \times 10^{58}$. The parameter that controls the approach to super-Poincare invariance in HST is $N^{-\frac{1}{2}}\sim \frac{1}{2} 10^{-29} $. This gives a gravitino mass of approximately $5 \times 10^{-2} $ eV, and a scale for superpartner masses within reach of the LHC. We will turn next to the quantum theory of dS space in the HST formalism.

\section{The end of the universe}

Observational data suggest that the universe has entered into a
phase dominated by a positive cosmological constant and will
approach de Sitter space in the relatively near future. We propose
that particle physics is well described by a model of eternal stable
dS space. The motivation is similar to that which held sway in the
early days of the superstring revolution. Although string theory is
a theory of gravity and should thus explain cosmology, it was
thought that the microphysics of particles should not depend on
early cosmological history and could be modeled by a theory in
asymptotically flat space-time. This project failed, because all
stable models in asymptotically flat space time turned out to be
exactly supersymmetric, while the world is, manifestly, not.

The Holographic theory of space-time provides an explanation of
this, which we sketched above. Lorentz invariance is conformal
invariance on the sphere at infinity, and one can then construct the
super-Poincare generators from the pixel variables and the CKSs.
Conversely, one may expect that since the dS holoscreen has finite
area, SUSY will be broken.

dS space is the Lorentzian continuation of the Euclidean 4-sphere.
Semi-classical study of dS space with radius $R$ reveals the
following properties

\begin{itemize}

\item dS space is a thermal system with a {\it unique} temperature
$T = \frac{1}{2\pi R}$ and an entropy $\pi (RM_P)^2$.

\item There is a maximum black hole mass in dS space. In general,
Schwarzschild-dS black holes have two horizons $R_{\pm}$ related by

$$ R^2 = R_-^2 + R_+^2 + R_+ R_- ,$$ $$ 2M =\frac{M_P^2}{R^2} [ R_+ R_- (R_+ +
R_- )].$$ The sum of the horizon entropies is always smaller than the
$M = 0$ case of empty dS space. When $R_- \ll R_+$, the entropy
deficit is $2\pi R M$.

\item Coleman-DeLuccia tunneling probabilities between two dS minima
satisfy the principle of detailed balance with entropies instead of
free energies. This indicates that the system is at infinite
temperature and that the entropy is the log of the number of
dimensions in its Hilbert space. For a subclass of
potentials\cite{abj} the transition probabilities from the lowest dS
minimum to negative c.c. Big Crunches are entropically suppressed.
They look like low entropy fluctuations of a finite system in
equilibrium at infinite temperature. Indeed, the maximal area causal
diamond in the crunching region has a microscopic area and that
region is plausibly modeled as a low entropy state.

\item Although the early and late time spatial slices in global
coordinates grow to arbitrarily large size, one cannot actually send
in small perturbations from the past with impunity. Most will cause
a Big Crunch before the dS throat is reached. Thus, these
perturbations should not be thought of as part of the theory of a
stable dS space. The entropy of perturbations that do have such an
interpretation is bounded by the entropy of dS space, because high
entropy initial states form black holes. This is the analog of the restriction of perturbations of AdS space to be normalizable. A major difference between these two systems is that the space of states in AdS is infinite dimensional despite the boundary conditions. One can, perhaps, make an analogy between the dS case and that of a compact phase space. Classically, there are an infinite number of solutions of the equations of motion on a compact phase space, but the Hilbert space of states is always finite dimensional. The difficulty in making this analogy precise
has to do with the fact that the nominal phase space is infinite dimensional. It is only the interpretation of the Bekenstein-Hawking formula as a micro-canonical entropy (the strong HP), which tells us that
dS space has a finite number of states.

\end{itemize}

All of these facts can be explained in a model with a sequence of Hamiltonians
$$H(r) =Z(r) P_0 + \frac{1}{N^2} {\rm Tr}\ f(\psi \psi^{\dagger}), $$ acting on a Hilbert space with dimension $e^{\pi
(RM_P)^2}$, and built from our fermionic pixel generators. $N= RM_P$. The polynomial $f(x)$ is order two and higher in the fermion bilinears. $P_0$ converges, as $R M_P \rightarrow \infty$ to the
Hamiltonian of a super-Poincare invariant theory of quantum gravity
in asymptotically flat space. The factor $Z(r) = \sqrt{1 - \frac{r^2}{R^2}}$, incorporates the redshift of particle energies on a trajectory of fixed static coordinate in dS space. We'll argue that the Hamiltonian puts the system into its maximally uncertain density matrix (after the usual ergodic time averaging), and that this is a finite temperature $T^{-1} = 2\pi R$, density matrix for the particle Hamiltonian $P_0$ of the (geodesic) observer at $r=0$. $Z(r)$ then incorporates the blue shift of the temperature as we approach the horizon.

By conventional large $N$ counting, the higher order part of $H$ has a spectrum bounded by
something of order $1/R$. We choose this part of the Hamiltonian to be chaotic\cite{berrydeutschsred}. This
means that for a generic initial state, time averaged expectation
values of many operators approach those in the maximally uncertain
density matrix in a relaxation time of order $R$. The Heisenberg
recurrence time for $H$ is $\sim e^{\pi (RM_P)^2} R$. For typical
measures on Hamiltonians on large Hilbert spaces, the chaotic properties
are true with probability one (any measure that is absolutely
continuous with respect to the Gaussian measure works).

A further, implicit, constraint on $H$ comes from the properties of
$P_0$. The bound on its spectrum is the Nariai black hole mass
$$||P_0|| = M_N = \frac{1}{2\sqrt{3}} R M_P^2 ,$$ and the entropy deficit
of its eigenspaces, relative to the full dS entropy is given by the
black hole mass formula
$$ \Delta S = \pi R_+ R_- ,$$ where
$$R^2 = R_+^2 + R_-^2 + R_+ R_- ,$$ and
$$ 2M R^2 = M_P^2 R_+ R_- (R_+ + R_-) . $$ For eigenvalues small
compared to $M_N$, this says that the maximally uncertain density
matrix is the thermal density matrix with temperature $T^{-1} = 2
\pi R$ for $P_0$. Thus, the thermal nature of dS space for a local observer is explained,
and the uniqueness of the dS temperature is a consequence of the fact that the underlying ensemble
is really the infinite temperature ensemble in a finite Hilbert space. It's only the fact that
$P_0$ does not act on large tensor factors of that Hilbert space, which makes the local temperature finite.

In addition, we postulate a commutator
$$[H, P_0] = M_P^2 g(P_0 / M_N ), $$ where $g(x) \sim x$ for $ x \ll
1$. This can be motivated by comparing the action of the generator
of static time translations on the cosmological horizon of dS space,
with the action of Poincare time translations on null infinity, to
which the cosmological horizon converges as $R \rightarrow\infty$. It's clear
that this equation is true as an order of magnitude estimate in powers of $N$. The fact that the leading order term is proportional to $P_0$ appears to be a constraint on the function $f(y)$.

If we take $P_0$ to be approximately the Poincare Hamiltonian for particle states, which we would
construct in taking the limit of vanishing c.c. as $N\rightarrow\infty$, we can verify that the entropy
-energy relation discussed above is satisfied for systems with a finite number of particles of total energy
less than the Planck mass.
These are described by block diagonal matrices with block sizes $\sum N_i \ll N$.
The entropy associated with the fermionic pixel generators in those blocks is particle entropy, and is ``unavailable" to
the horizon. Moreover, in order for the particles to decouple from the horizon, we must also freeze the generators which connect the particle variables to the horizon in the matrix $\psi_i^A$\cite{holounruh}. The number of these generators is $2N \sum N_i$. Thus the number of free horizon generators is, for large $N$, approximately $$(N^2 - 2N \sum N_i ),$$ and the entropy deficit is proportional to $N\sum N_i \propto P_0 / T$. The particle energy is proportional to $\sum N_i$ and the dS temperature is proportional
to $N^{-1}$, so the entropy deficit scales like that of a thermal ensemble at the dS temperature. A description of how the entropy energy relation works out for black hole states in dS space may be found in \cite{bfm}.

The second term in $H(r)$ has a highly degenerate spectrum, with level spacing $\frac{1}{R} e^{ - \pi (RM_P)^2} $, as a consequence of which much
of the dynamics it describes takes place on extremely long time
scales. The shortest scale over which the higher order term is active
is of order $R$. $P_0$ on the other
hand contains all of timescales of local physics. Furthermore, the
equation for the commutator tells us that $P_0$ eigenspaces of
energy $\ll M_N$ are stable under $H$ evolution for a very long
time. The $P_0$ eigenstates will eventually decay back to the dS
vacuum, but the time scale for that is of order $R$ or greater.
These equations do not take into account the effect of other
conservation laws. For example, an electron bound to the observer
at the origin cannot disappear until a positron spontaneously
nucleates in its vicinity (with a compensating negative charge
density on the horizon). The probability for this is $~ e^{ - 2\pi
R m_e}$, so this particular $P_0$ eigenstate is much more stable
than others.

Below, we will provide a more general context for understanding the relation between
$H(r)$ and $P_0$, in terms of observers following trajectories of varying acceleration, within the
same causal diamond. A particular set of such observers in the maximal causal diamond of dS space
are the time-like trajectories with fixed spatial coordinates in the static coordinate system. We have in fact written the general Hamiltonian
along such a trajectory by incorporating the red shift factor $Z(r)$. $Z(0) = 1$, for the geodesic passing through the origin, and $Z$ decreases
to $o(\frac{1}{RM_P})$ for the maximally accelerated trajectory, corresponding to the ``stretched horizon". The stretched horizon
Hamiltonian is
an operator with a chaotic spectrum of eigenvalues spread between $0$ and $1/R$ with density $e^{ - \pi (RM_P)^2}$.

The instability of generic $P_0$ eigenstates is illustrated by the
black hole formula, with $M$ interpreted as the $P_0$ eigenvalue.
For small black holes, we see that the black hole is a low entropy
state whose entropy {\it decreases} as its mass {\it increases}.
Hawking's formulas imply that it decays down to a Planck scale
remnant in a time of order $\frac{M^3}{M_P^4}$, and it is likely
that that remnant decays unless it carries a charge. Charged states
bound to the observer will be much more stable, but will decay when
a thermal fluctuation nucleates an opposite charge at the position
of the first. Note that the simplest instability of all is that of
arbitrary localized excitations, which are not bound to the
observer. These simply travel out to the horizon in a time of order
$R$, and become part of the thermal bath. From the observer's point
of view this is a decay back to the dS vacuum.

These estimates show that the dS recurrence time has nothing to do
with ordinary physics. It is a property of the thermal bath on the
horizon, but all localized excitations dissipate on a much shorter
time scale and the Hamiltonian $P_0$ has nothing to act on. The dS
recurrence time is only relevant to an attempt to explain the low
entropy beginning of the universe as a thermal
fluctuation\cite{dks}. This idea, originally due to Boltzmann, is
wrong, which is the main point of \cite{dks}. In the next section, we will attempt to provide the
beginnings of an alternate explanation for the fact that the
universe began in what appears to be an improbable state.

\subsection{Accelerated Observers and Fast Scramblers}

We end this section by noting that we can put
the relation between the Hamiltonians $H(r)$ in a more general context. Two space-time points with time-like separation determine a causal diamond. There is a unique time-like geodesic connecting them. However, there are also an infinite number of accelerated ``Unruh trajectories", which have the same causal diamond. Let us consider what to expect for Unruh observers in Minkowski space. For simplicity, assume we're talking about a large causal diamond in asymptotically flat space-time. QFT leads us to expect a higher temperature for Unruh observers with larger accceleration. A simple way to get a qualitative version of the relation between Hamiltonians for observers with different accelerations can be constructed using our construction of particle states in terms of matrices. The ``algebra of functions on the holographic screen of dS space" is the tensor product of the algebra of $N \times N$ matrices ($N \sim RM_P$) with a matrix algebra ${\cal M}_I$ acting on the internal spinor bundle with Dirac cut-off. Particles are described by restricting this algebra to a direct sum of block diagonal algebras of the form $\sum (N_i \times N_i ) \otimes {\cal M}_I^i $, and restricting attention to the subset of quantum pixel operators with fermionic indices belonging to this sub-algebra of matrices. Supersymmetry tells us that the particle Hamiltonian is quadratic in these fermions, and if we write a quadratic form of the type ${\rm Tr} \psi^{\dagger} \psi$, then it does not couple the particle variables to the off diagonal matrices. Higher order terms do couple all of the $\sim N^2$ fermionic variables together. Susskind and Sekhino\cite{fastscramble} have argued that general Hamiltonians that are functions of traces of large $N$ matrix variables are {\it fast scramblers, i.e.} systems that equilibrate rapidly because all variables are coupled together. Thus, a sequence of Hamiltonians with an acceleration dependent quadratic term in fermions, plus a collection of higher order polynomial traces, could model the one parameter family of Unruh observers. We propose that the Hamiltonian describing the last Planck time of evolution in a causal diamond whose radius in Planck units is $N$, has the form
$$H(a) = Z(a) P_0 + \frac {1}{N^2} {\rm Tr}\ P(\psi \psi^{\dagger} ) ,$$ where $Z(a)$ is a redshift factor. It is equal to $1$ for the geodesic trajectory and approaches $1/N$ for the maximally accelerated stretched horizon observer.
$Z$ incorporates the gravitational redshift of particle energies. With these definitions, the geodesic observer will see approximately stable particle eigenstates, while the accelerated observers will encounter a heat bath of increasing temperature, due to their stronger coupling to the non-particle like degrees of freedom in the diamond. This idea will be pursued in a future paper\cite{holounruh}.

To summarize, we have sketched a quantum theory of stable dS space, which becomes super-Poincare invariant as the dS radius goes to infinity. We've isolated localized particle observables from the full set of degrees of freedom of the system, and noted that the dS temperature arises because the localized states with non-zero energy have less entropy than a typical state in the dS vacuum ensemble, which is the maximally uncertain density matrix. This observation will be the basis for understanding the origin of the second law of thermodynamics in cosmology.

\section{The beginning of time}

There is nothing in the principles of quantum mechanics that forbids
us to discuss a system that began a finite time in the past, as long
as the Hamiltonian is time dependent. We've already argued that time dependence of the
Hamiltonian is in fact required if we are to build the concept of
{\it particle horizon}, and causality more generally, into a holographic model of cosmology. For cosmology with four non-compact dimensions, the
Hilbert space of each observer is a nested sequence ${\cal H} (N, x)
= \bigotimes{\cal P}^{N(N+1)} ,$ where the product is the tensor product and $N$ is a positive integer. We begin
with a model in which $N$ ranges from $1$ to infinity. $x$ labels a
point on a cubic lattice, though a much more general choice of 3 dimensional simplicial complex gives
the same coarse grained results. The lattice describes the topology of the Big Bang hyper-surface in space-time. $x$ is a label for the quantum version of a congruence of time-like trajectories. With each such trajectory we associate a sequence of unitary operators, which describe time development according to proper time along that trajectory. The nested structure of the Hilbert space corresponds to larger and larger intervals of proper time, and the dimension of each Hilbert space tells us the area of the holographic screen of the causal diamond for that interval.
Overlap rules, which we will describe below, knit the description of physics according to each of these ``observers", into a coherent space-time picture.
In cosmological space-times, it is natural to take all the causal diamonds to begin on the Big Bang hypersurface, which in HST is defined by the statement that the observer's Hilbert space has the smallest possible dimension $D_{\cal P} \equiv e^{\cal L}$. The Big Bang is thus non-singular, but in no way approximately described by QFT.

Causality is insured by assuming a sequence of unitary operators
$U(k, x) = V(k,x) W(k,x)$, where $V(k)$ acts in ${\cal H} (k,x)$ and
$W(k)$ in its tensor complement in ${\cal H} (N_{max} (x) , x)$ \footnote{Here we've allowed for the possibility that $N$ has a finite bound, which might be $x$ dependent. In two of the models we present, it is $x$ independent, while the third has such dependence.}. The
states in ${\cal H} (k, x)$ are said to be ``inside the particle
horizon at the time labeled by $k$'' . The time dependent
Hamiltonian is defined by $U(k,x)U^{\dagger} (k - 1,x) = e^{ - i
H(k)} ,$ where we will insist that we have the same Hamiltonian for
each $x$. In fact, given the choice we will make for the
Hamiltonian of a single observer, the Hamiltonian must be $x$
independent in order to satisfy the consistency conditions.

We choose $H(k)$ at each $k$ from a random distribution, with the
constraint that for large $k$ it approaches $ u^{\dagger}
(k) (h(k) + P(k)) u(k) $. Here $h(k)$ is the Hamiltonian of a $1+1$
dimensional conformal field theory, with a central charge proportional
to $k^2$ and cutoff momentum $\frac{1}{k}$ on an interval of size $k$, in
Planck units. $P(k)$ is a random irrelevant perturbation of $h(k)$
and $u(k)$ a random unitary in ${\cal H} (k, x)$. The choice of central charge
will be justified below.

We define the
overlap rule to be
$${\cal O} (N; x,y) = {\cal P}^{N(x,y)(N(x,y) + 1)} ,$$ where $N(x,y) = N - d(x,y)$ and $d(x,y)$ is the
minimum number of lattice steps between the two points. On a cubic
lattice, the locus of all points at fixed $d(x,y)$ from a given $x$
is a cube tilted at $45$ degrees to each axis. However, in
space-time we measure distance in terms of causality. All points on
this cube are at the same spatial distance. For large $k$ the
space-time geometry becomes spherical. The dynamical consistency
conditions are all satisfied and we see the emergence of a
homogeneous isotropic cosmology. It is exactly homogeneous, because the consistency conditions
require us to choose the same initial density matrix for all $x$. Isotropy sets in for large $k$ because in a general lattice with the topology of flat three dimensional space, the locus of all points that are a fixed number of steps from a given point becomes a sphere, according to the distance defined by our emergent Lorentzian geometry, when the number of steps is large. We impose isotropy on the description of an individual observer by
insisting on Hamiltonians that are invariant under the $SO(3)$ rotations on the spinor bundle variables $\psi_{i\ P}^A$.

Our choice of Hamiltonians was motivated by saturating the covariant
entropy bound. Time evolution is a sequence of random unitaries, which will sweep out the whole Hilbert space at each $k$.
More precisely, the sequence of states swept out in our discrete time evolution is a random walk on the space of states of the system. In FRW space-time one can only saturate the CEB for flat
spatial sections with equation of state $p = \rho$\cite{fs}. Another
indication that we have flat spatial sections is that we have a
scale invariant system. Flat FRW metrics with single component
equations of state all have a conformal Killing vector, but
negatively curved sections break the scale invariance\footnote{We
have ruled out positively curved spatial sections by our choice of
spatial topology, but it is also obvious that our model does not
suffer a Big Crunch, because the maximal area causal diamond of an observer in a Big Bang $\rightarrow$ Big Crunch cosmology has finite area.}.

In keeping with the spherical nature of the particle horizon, we describe our system in terms of the variables introduced above,
$$[(\psi)_{i\ P}^A , (\psi^{\dagger})_B^{j\ Q} ]_+ = \delta_i^j \delta^A_B Z^{PQ}.$$ The integer $N$ is the size of the matrices. Note that we have chosen
to keep the pixel Hilbert space, which is determined by the non-commutative spinor geometry of the internal dimensions, fixed as the particle horizon increases. It might be that there are
more general cosmologies in which this rule is not enforced, but the discrete nature of changes to the non-commutative geometry makes this a completely consistent and stable choice.
If we were thinking in terms of the traditional picture of moduli of the internal geometry, perhaps with a potential on moduli space, we would assume that at least in the very early universe, the moduli must vary with time. The scaling of the number of degrees of freedom with $N$ suggests a central charge $\sim N^2$, for the CFT. Perhaps the relevant CFT is related to the Kac-Moody super-algebra associated to the pixel algebra.

A change of $N$ by one, corresponds to a change in entropy of order $N$, so $N$ is simply proportional to cosmological time in Planck units. Following \cite{holocosmbfm} we can verify that the space time energy and entropy densities, defined by the $1 + 1$ CFT energy and entropy, divided by the spatial volume $\sim N^3$ satisfy
$$\rho \sim \frac{1}{N^2}\ \ \ \ \ \sigma\sim \frac{1}{N} ,$$ as expected for a flat FRW universe with $p = \rho$. Moreover, our choice of an overlap rule {\it quadratic} in $N$ solves the problem with the entropy of overlaps, which was encountered in \cite{holocosmbfm}. Note that, without violating the rules of quantum mechanics, we can take a finer and finer time slicing as $N$ gets large, changing the entropy only by $O(1)$ at each step. However, there does not seem to be a way to do this without violating rotation invariance. We can only add fractions of angular momentum multiplets. A heuristic way to understand this is that such tiny time steps only add one pixel of information, and given the intrinsic cutoff of our non-commutative geometry, there is no way to add a single pixel in a rotation invariant way.

The average total energy in the ensemble of states of the CFT scales like $N$, and the entropy like $N^2$, which are of order the mass and entropy of a black hole filling the horizon volume.
Thus the model fits a heuristic picture of the universe as a closely packed collection of black holes, which interact and coalesce in such a way as to fill the horizon volume at every moment of time. It's easy to see that the such a Dense Black Hole Fluid (DBHF), would indeed behave like a $p = \rho$ fluid \cite{bf1}, and we consider our model to be a mathematical definition of what the words DBHF mean. We've found that listeners get hung up on this heuristic picture and try to explain
our results to themselves in terms of it. This is, we believe, a bit confusing and we've begun to regret the terminology. Our model is well defined mathematically and has the thermodynamics properties we attribute to it. Trying to ``explain" the pressure in terms of some intuition of how a solution of GR with many black holes separated by distances of order their Schwarzschild radius ``should" behave is not necessarily a useful exercise.

\subsection{Holographic Eternal Inflation (HEI)}

There is a variant on the DBHF cosmology, which describes a
transition to something like an eternally inflating universe, but one quite different
than QFT based models for eternal inflation. One simply stops the growth of the
Hilbert spaces at some fixed value $N_{max}= N$, and makes a transition
to the fixed Hamiltonian $H_{dS}$, which we have used to describe a
stable dS space with entropy $N_{max} {\cal L}$.
Indeed, it might well be
that the DBHF Hamiltonian $H(N_{max})$ is, up to a rescaling, already a suitable candidate for
$H$. Recall that the spectrum of $H_{dS}$ had all the characteristics of
a random Hamiltonian, consistent with the existence of the Poincare
Hamiltonian $P_0$. The largest eigenvalue of $H_{dS}$ is of order $1/N$, while typical eigenvalues
of $H(N)$ are $\sim N$.

We definitely expect some kind of rescaling of time relating the DBHF description and that of the terminal static dS space. We've argued that the random Hamiltonian describing dS space is that of an accelerated observer on a holographic screen just inside the cosmological horizon. On the other hand, the DBHF Hamiltonian was constructed to model evolution in ordinary cosmological time, which is the time of a geodesic observer. This is somewhat problematic conceptually, since we view the particle horizon of the DBHF as filled with a black hole. The only reasonable geodesic observer one can imagine, is one following the last stable orbit around the black hole. This is at a distance of order one Schwarzschild radius from the horizon, so energy measured by such an observer would scale like the black hole mass in asymptotically flat space. Energy of order $1/N$ corresponds to a redshift that would be encountered on a stretched horizon whose {\it area} differed from that of the horizon by one Planck unit. In the next subsection, we will describe a stable dS space, embedded in a marginally trapped surface in the $p=\rho$ cosmology. One should view the geometrical picture of the relation between the two Hamiltonians $H(N)$ and $H$ in the context of that geometry. It has a horizon with a $p=\rho$ FRW on the outside and dS space on the inside. In the context of our model, we go from $H(N)$ to $H$ by performing a dilatation of the $1+1$ CFT, obtaining a Hamiltonian on an interval of order $N^{3}$ with momentum cutoff of order $N^{- 3}$, with the same central charge.

We will call the quantum cosmology defined by these rules, the {\it holographic eternal inflation} or HEI model. In order to define a consistent HST, we must modify the overlap rules after we reach the maximum size Hilbert space. Two points that are separated by more than $N_{max}$ steps on the lattice never have any overlap. This means that the HEI model describes a system with an infinite number of physical states. Take a random point on the lattice and draw the tilted cube of all points that have overlap with it. Now tile the lattice with non-overlapping tilted cubes of size $N_{max}$, centered on a collection of points, no pair of which have any overlap with each other. We can view the Hilbert spaces associated with these centers as independent degrees of freedom with no constraints on their dynamics. Note however that the Hamiltonian and initial state of these independent systems is the same, as a consequence of the overlap conditions with all the non-central points in the tilted cubes. In this model, we can just as well state that the independent degrees of freedom at the centers of different tilted cubes are just gauge copies of the degrees of freedom in a single cube. Below, we will introduce a different model, which treats these degrees of freedom as physical. This is the key to a theory of inflation.

\subsection{Emergent effective field theory}

The HEI cosmology fits all of the coarse grained properties of a
generic classical solution of the Lagrangian

$${\cal L} = \sqrt{-g} [R - \frac{1}{2} g^{\mu\nu}
\partial_{\mu}\phi \partial_{\nu} \phi - \Lambda ],$$ subject to
the non-holonomic constraint
$$ g^{\mu\nu}\partial_{\mu}\phi \partial_{\nu}
\phi < 0 .$$ The constraint fixes the gradient of $\phi$ to be
timelike, which implies that we can introduce an FRW coordinate
system in which $\phi$ depends only on time. Note that this low
energy field theory description {\it cannot} describe most of the
microscopic quantum states of the model. Like the near horizon
states of a black hole, these live on the particle horizon, and
cannot be modeled in terms of bulk fluctuations of the fields. The
non-holonomic constraint eliminates solutions of the classical field
theory, which have nothing to do with the quantum system it is
modeling. This is an illustration of what might be called {\it
The Jacobson-Padmanabhan Principle}: Einstein's equations are the laws of
thermodynamics applied to a system whose entropy is measured by the
area of Rindler horizons. It does not follow that the quantum
mechanics of the system is obtained by naive quantization of those
equations. For certain background space-times quantization of
Einstein's equations is a valid procedure in the semi-classical
approximation, but not for all. We {\it always} expect naive quantization of the
low energy field theory to fail when we examine states that saturate the CEB.

Indeed, we've seen how particle physics can arise from the HST formalism when the Hamiltonian
for a region of size $N$, decouples the $N^{3/2}$ particle-like degrees of freedom from the full matrix of $N^2$ variables.
The CEB is saturated precisely when all these variables are coupled together. In HST {\it quantum} field theory arises as an approximate description of particle physics, valid when ``particles decouple from black holes". On the other hand, Jacobson's arguments suggest that the {\it classical } Einstein equations are just an encoding of the thermodynamics of any system
which has an emergent space-time and obeys the Bekenstein-Hawking law. Since the space-time of the HST model is NOT
required to satisfy the vacuum Einstein equations, the low energy field theory description of its thermodynamics will always need extra field degrees of freedom. These are just macroscopic variables, and it makes no more sense to quantize them than it does to quantize hydrodynamics or the metric itself.

It's instructive to compare this effective field theory, with those that arise from string models on non-compact space-times with c.c. $\leq 0$ . There, string theory supplies us with a set of boundary correlation functions, and we introduce fields to provide an approximate calculation of those correlation functions. It is wrong to quantize those bulk fields. They introduce too many degrees of freedom into finite regions of space-time. This leads to uncontrollable UV divergences and to an incorrect description of black holes. String field theory eliminates the UV divergences to all orders in string perturbation theory, if we are dealing with an exactly supersymmetric model in $\geq 4$ asymptotically flat dimensions, but the string field Lagrangian is not well defined. It must be corrected at each order in perturbation theory and the result is a non-Borel summable series of corrections. It does not get black hole physics right. The correct procedure for dealing with effective fields in asymptotically flat space is to calculate the S-matrix for stable particles, and find an effective action, which reproduces it order by order in the momentum expansion. Ancient theorems of S-matrix theory assure us that this can be done for theories of massive particles, but we don't even know that it is true when massless loops are included in the calculation. More seriously, it seems highly unlikely that the full S-matrix that includes production and decay of a meta-stable black hole, can ever be encoded in such an effective action.

Our point is that the fields we introduce in discussions of approximations to models of quantum gravity, are highly constrained objects, which should not be allowed the conceptual weight we attribute to them in quantum field theory. If effective field theory fails for processes involving production of black holes in asymptotically flat space, it must be even more constrained when applied to situations where black-hole like states are typical, rather than restricted to certain controllable kinematic regimes. This is in fact our claim about the effective field theory of the HEI space-time. The underlying quantum model is exactly homogeneous, and is isotropic in the emergent space-time limit. The effective field theory describes {\it only} the classical homogeneous hydrodynamics of the quantum system. The true quantum variables do not have even an approximate field theory description. The reason that there is a more detailed correspondence between bulk field theory and the quantum theories of asymptotically flat space-time is that the approximate and exact theories use the same description of the Hilbert space of scattering states, and that by its nature, scattering is a process where black hole production can be suppressed by appropriate choice of kinematic variables.
In AdS space\cite{BDHM} the Fock space structure begins to break down at finite order in the $1/N$ expansion. The derivation of effective field theory from boundary CFT is still incomplete\cite{joeetal} beyond the semi-classical approximation.

\subsection{Stable dS space as a black hole in the DBHF}

If $\Lambda \neq 0$, the holographic inflationary FRW (HEI) model is massively
redundant. It introduces an infinite number of observers to describe
a physical system where only a finite number of measurements are
possible during the entire history of the universe. We will use this redundancy to
good effect below, when we construct a model of inflation, but it is {\it not} the correct
description of a universe which begins as the DBHF and evolves to a stable dS space.

A better model can be constructed by restricting the $\Lambda \neq 0 $ region to a
finite tilted cube of the lattice, and using the $\Lambda = 0$
rules far outside that. On the boundary of the cube, we choose a Hilbert space whose
dimension is determined by the entropy of dS space, and assign it the random sequence of Hamiltonians of
the DBHF $\rightarrow$ dS FRW model of the previous subsection. For any point outside the cube, we declare that the overlap with all points inside the cube is empty, while the overlap between each point and any point on the boundary follows the DBHF rules but saturates when the overlap is the size of the boundary Hilbert space. On the interior, our choice of Hamiltonians and overlap rules depends on a choice of coordinate system. At early times we have co-moving observers in the $p=\rho$ FRW universe, but we might want these observers to asymptote in the future to observers with a fixed position in the static dS coordinates. The overlap rules for such observers can, in principle, be read off the geometry, but the required calculations are non-trivial. They are not the same as the rules for the DBHF $\rightarrow$ dS FRW model of the previous subsection. In order to make the HST rule that one step on the lattice of observers corresponds to an overlap Hilbert space that is missing exactly one pixel, compatible with the total dimension of the Hilbert space in the interior, we must choose the number of steps between the central point of the tilted cube and its boundary, equal to the radius, $R$, of dS space, in Planck units.

Inside the tilted cube we must find, at each point, a set of time dependent Hamiltonians whose initial member couples together only degrees of freedom in a tensor factor of dimension $e^{\pi n^2}$, and eventually operates freely on the full Hilbert space of dimension $e^{\pi N^2}$\cite{tbwfds}. $n$ and $N$ are , respectively, the inflationary Hubble radius, and the Hubble radius corresponding to the observed c.c. . The Hamiltonians at different points at the interior of the cube, and the choices of initial state, must satisfy the overlap rules of HST.
Geometric intuition suggests that there must be {\it many} ways to do this, corresponding to different congruences of time-like trajectories within the causal diamond of a given observer. Since we have implicitly chosen the tilted cube to represent the horizon volume of a particular geodesic observer in a future asymptotically dS space-time, we choose the Hamiltonian at the center to converge to the Hamiltonian, $H(r = 0)$ of that observer. We will call the time dependent sequence of Hamiltonians $H(0 ,t)\rightarrow H(0)$. This Hamiltonian {\it approximately} decouples the full system into a tensor product of Hilbert spaces of dimension $\sim e^{c N^{3/2}}$, representing localized particle excitations (and black holes of radius $\ll N$) in individual horizon volumes. The terms coupling these different Hilbert spaces are of order $1/N$. Note that the total number of
horizon volumes scales like $N^{1/2} \ll N^3$, the number of points in the interior of the cube.

Depending on what we want to describe, our choice of Hamiltonians at the $N^3$ interior points will be different. For the moment, we describe a system of maximal entropy, whose interior hydrodynamics evolves (approximately) homogeneously from the DBHF to dS space. A natural coordinate system is one which evolves from the FRW coordinates of the DBHF to the static coordinates in dS\cite{dks}. We assign points $k$ steps from the center of the cube to fixed positions $r_k$, in the static coordinates, with the change in $r_k - r_{k-1}$ determined by the overlap condition that the largest causal diamond in the intersection between the diamonds at $r_k$ and $r_{k-1}$ has an area in Planck units of one pixel, $4 {\rm ln}\ {\cal P}$. The area of the intersecting causal diamond can be computed at any time and for any pair of trajectories labeled by their final static coordinates in the coordinate system of \cite{dks}.
The full FRW geometry is given by
$$ds^2 = - dt^2 + c\ \sinh^{2/3} (3t/R) d {\bf x}^2 ,$$ in coordinates (not those of \cite{dks}) that approach the flat slicing of dS space in the future\footnote{Our geometry is always expanding and the flat slicing covers all of the expanding part of the dS manifold.} . The constant $c$ is a function of the ratio between the asymptotic c.c. and the coefficient of $\frac{1}{a^6}$ in the energy density of the early $p=\rho$ universe. Going to the coordinate system of \cite{dks} and computing all of the overlap areas, is a straightforward but tedious exercise. Given its results, we can write down the overlap rules for the Hilbert spaces at arbitrary pairs of points on the lattice. One must then invent a dynamics, which satisfies the overlap constraints of HST, reduces to $P_0 (t)$ for the central point, and becomes a random Hamiltonian on the full Hilbert space of dimension $e^{\pi N^2}$ as we approach the boundary. This will, by construction, be consistent with the overlap rules for the exterior.

We believe, but have not verified that the correct sequence of Hamiltonians is given by

$$H(r, n) = \sqrt{1 - \frac{r^2}{R^2}} P_0 (n) + \frac{1}{n} V(n) ,$$ where $P_0 (n)$ is the approximation to the Poincare Hamiltonian, which appears in the Super-Poincare algebra when we restrict the spinor bundle over the two sphere to be just $n \times n+1$ matrices. $V(n)$ is a fast scrambler Hamiltonian, constructed as a trace of operators quadratic and higher in the fermion bilinears $(\psi^{\dagger} \psi )_i^j $, and normalized to be $O(1)$ in the large $n$ limit.
This proposal will be explored further in \cite{holounruh}.

These rules, by construction, give rise to a geometry with a marginally trapped spherical surface of area $\pi R^2$ in Planck units. Far from that surface the dynamics of individual observers approaches that of the DBHF. Inside the surface, at late times, we have the model of static dS space.

The coarse grained
description of this quantum system is a black hole in the $p=\rho$
universe, whose interior is filled with a horizon volume of the
DBHF-dS cosmology described above. In \cite{holocosm} we pointed out that the Israel junction conditions
are satisfied by such a configuration. The coarse grained description of the resulting cosmology can be described by an effective theory of
a scalar field coupled to gravity, with a potential $V(\phi)$ and the non-holonomic constraints that the angular momentum
vanishes, and the horizon area is fixed to be $4\pi R^2$ . The potential has a local minimum at the origin with $V(0) = 3 R^{-2}
$, and falls rapidly to zero as $|\phi|\rightarrow\infty$.
The initial conditions, on a time slice $t_I$ a bit after the (singular) Big Bang, are
$$\phi (0, t_I) = 0 , \phi (r , t_I) \rightarrow\phi_0 + e^{-r/R} , $$
$$\dot{\phi} (r, t_I) = v \gg \sqrt{2 V(0)} .$$ The potential is chosen so that the field in the region $r < R$ remains within the basin of
attraction of the origin. $\phi_0$ is chosen so that $V(\phi_0) \ll V(0)$. It's clear that the exterior region will collapse to a black hole embedded in the $p = \rho$ FRW space-time. We tune the potential and the initial conditions so that this black hole has horizon radius $R$. Inside the horizon, the positive c.c. provides the possibility of avoiding a singularity. Again, we choose the initial conditions and the potential so that the interior approaches the static patch of dS space over most of its volume. The Israel junction conditions assure us that there should be non-singular solutions of this form. The dS space is {\it not} separated from the exterior by a singular Einstein-Rosen bridge.

Indeed, at very late times, solutions of this sort can be extracted from the work of Mazur and Mottola\cite{mazmot}. These authors constructed static solutions of Einstein's equations, corresponding to an asymptotically flat Schwarzschild solution separated from a dS interior by a shell of static matter with equation of state $p=\rho$. They emphasize that their solutions have no horizon. Our model corresponds to the thin shell limit of their solution, which does have a horizon.

We would like to emphasize the way in which this effective field theory description, and particularly the field $\phi$, arises. We first construct a consistent HST model using the fundamental pixel variables, and use the HST relation between quantum mechanics and space-time geometry to abstract a space-time geometry as the coarse grained description of this quantum model. There is no reason for this geometry to satisfy the vacuum Einstein equations. {\it Thus, in order to model the geometry in effective field theory we must INVENT non-gravitational degrees of freedom to give rise to the non-zero Einstein tensor of the geometry.} This is very different from the conventional approach to ``string cosmology", where we start from fields that are ``in string theory"\footnote{Which is to say they are part of the effective field theory treatment of string theory in flat space.}, and use them as inflatons. As we've emphasized above, the effective field theory description of asymptotically flat models of gravity is appropriate for situations in which, outside of a sufficiently large causal diamond, the particle degrees of freedom in HST decouple from the much more numerous set of variables that describe a black hole of area equal to that of the holoscreen of the diamond. This is simply not the case for the dS black hole in the $p=\rho$ space-time. In that space-time, the CEB is saturated everywhere in space-time and field theory can describe only the coarse grained hydrodynamics of that model, as anticipated by Jacobson.

The relation of this space-time to particle physics is more complicated. At the central point of the interior dS region, we have for a causal diamond of area $\pi n^2 L_P^2 $,
$$H(n) = P_0 (n) + \frac{1}{n} V(n),$$ For times less that $R = NL_P$, but $n \gg 1$, the $P_0$ evolution is active, and the fast scrambling is negligible. As times become larger than $R$, the Hamiltonian is a constant and we cannot neglect the thermalization of the particle degrees of freedom with the horizon. More and more particle states begin to "decay" into the vacuum, as localized excitations exit the static observer's horizon. Effective field theory of the particle degrees of freedom is a valid tool for studying particle physics over the time scales preceding decay, as long as the kinematic conditions do not lead to the creation of black holes whose size grows with $R$. It has nothing to do with most of the degrees of freedom on the horizon, which are responsible for the stable equilibrium with the DBHF in the exterior.

However, in the model of cosmology described so far, there is no time at which these particle degrees of freedom decouple from the rest of the system. The initial state is generated by the DBHF and is totally scrambled. Thus, although the Hamiltonian of the central observer coincides with that of a system of particles in dS space, the overwhelming probability is that the state of the system in the dS era belongs to the dS vacuum ensemble. The geodesic observer actually sees nothing, except an occasional thermal fluctuation, which produces a particle or two, and much less probable fluctuations in which Boltzmann's brain, and Einstein's, spontaneously pop into existence for fractions of a second, before their internal pressure blows them apart into the vacuum. Thus, following an argument that has by now become tiresome, even if we condition our probabilities on the existence of conscious observers, this model is doubly exponentially unlikely to exhibit the features of the world we know.

The model of a universe which evolves from the DBHF to a black hole with dS interior embedded in the $p=\rho$ universe has the coarse grained features of what we think the history of our universe looks like. It begins in a non-singular manner, goes through a period that looks like the $p=\rho$ Big Bang universe, and asymptotes to dS space with a fixed c.c. It is automatically homogeneous, isotropic and flat, with no fine tuning of the initial state. Indeed, this model is everywhere and always in equilibrium, with the maximal entropy it can have at any time. What it is lacking, in order to resemble what we see, is localized excitations, at any period in its history. We will sketch a model for the production of those excitations, in the next section.

\section{Heuristic Model of the Universe}

None of the models described so far is a realistic description of the universe we
observe. We will now present an idea for constructing a realistic
model along these lines. It does not correspond to a complete set of
overlap rules and Hamiltonians satisfying the rules of holographic
cosmology, but it is much closer to that goal than previous
attempts\cite{holocosm}. The entire infinite lattice evolves
according to the DBHF rules until the coarse grained system
resembles a $p = \rho$ FRW. Most points will continue with those
rules, apart from a set of tilted hypercubes that evolve to dS
universes embedded as black holes in the DBHF, as above. In the absence of a complete
quantum model, we will proceed to use Einstein's equations as a guide to how the system evolves after these black holes
form.

Black holes in an expanding universe can either remain isolated or collide. In a ``normal", mostly empty, universe, they could also evaporate, but this is not true if the ambient FRW is the DBHF. The black hole cannot increase the entropy of the universe by evaporating, and so a single black hole is exactly stable. Colliding black holes will form larger black holes. The interior dynamics of such solutions might exhibit some inhomogeneities before evolving into an empty dS space, but we have not been able to think of a mechanism for obtaining small curvature fluctuations from such collisions. Instead it seems most plausible that the interiors will be black holes in a dS universe, which will slowly evaporate back to the dS vacuum. It is very unlikely that any kind of organized life could evolve in such a universe.

The picture of isolated black holes in a $p = \rho$ universe, with a dS interior, and a variety of values for the c.c., seems ripe for the application of anthropic selection criteria. However, in order to make this plausible, we have to find a model in which some kind of life is possible. In order to do this, we must a construct a model universe similar to the dS black hole in the DBHF, but which has a natural mechanism for generating a set of localized excitations of the asymptotic Poincare Hamiltonian, $P_0$, which can lead to the evolution of life forms which are not Boltzmann brains. If possible, we would like to do this while using as little input as possible about any particular form of biology. Biology has not yet been, and may never be, derived from underlying physical principles (simply because of the complexity of biological systems) and we have no way of knowing how probable our form of intelligent life is, and whether other kinds ({\it e.g.} Fred Hoyle's Black Cloud or the Jovians of Hartle and Srednicki) are more probable, let alone what is possible with a low energy effective field theory different from the standard model. We might one day be able to address the question of whether HST allows other LEQFTs than the standard model in a universe with a small positive c.c. , but we are unlikely to have answers to more detailed biological questions in the foreseeable future. We will try to restrict our anthropic speculation to simple physical criteria like the existence of localized entropy producing structures, well modeled by QFT, which are surely necessary for the existence of any conceivable complex system that we could imagine calling {\it life}. We will show that, given the tools we have in hand, the most plausible model will be one in which we combine together multiple horizon volumes of the HEI model, with c.c. $\Lambda_I$, and embed them in the DBHF $\rightarrow$ dS black hole model, where the dS space has c.c. $\Lambda \ll \Lambda_I$.

Ironically, in this model we will find that most of the traditional roles of inflation are taken by other actors, while the real raison d'etre of inflation will be only to generate localized fluctuations. Indeed, we have already seen that homogeneity, isotropy and flatness, follow for all of our models, starting from generic initial states\footnote{Of course, for the dS black hole in the DBHF homogeneity and flatness are only approximate, for small c.c.}.

\subsection{The holographic theory of inflation and fluctuations}

The model of a universe whose coarse grained description is a black
hole embedded in the DBHF, with de Sitter interior, resembles our
own universe in a very broad brush manner, but each observer is in a
maximum entropy state (consistent with the CEB) at all times. This
follows from the microscopic mathematical model we have sketched, as
well as from the the CEB applied to the coarse grained space-time
geometries that follow from it.

We now outline a more realistic picture of the universe in which
we invoke a temporary period of inflation, with a c.c. $\Lambda_I$
much larger than the c.c. $\Lambda$ that describes the late time
history of the universe. The model contains degrees of freedom
modeling the full entropy of multiple horizon volumes of the
``inflationary dS space", but only one copy of the final dS horizon
volume. The relation between the two is given by the rough formula

$$e^{3 N_e} \sim \frac{\Lambda_I}{\Lambda} = (\frac{R}{R_I})^2 .$$

This is motivated by the following idea. We consider the HEI model at the interior points of the cube. Starting from the central point, we consider a tilted cube consisting of points $\leq R_I M_P \equiv n$ steps from the center.
The overlap rules of this model say that all the degrees of freedom at all of those points are visible to the observer at the central point. Apart from boundary effects, which are negligible when $n \ll N \equiv RM_P$, we can tile the tilted cube of $N$ steps with tilted cubes of $n$ steps.
The basic strategy of the yet to be constructed model is to define new overlap rules between the trajectories in {\it all} of the small cubes, which allow the Hilbert space at each of those points to grow to dimension $e^{N(N+1) {\cal L}}$ \footnote{Of course what we mean here is that we think of each point having the full Hilbert space at all times, but the Hamiltonian couples together more and more degrees of freedom, starting from the static Hamiltonian $H_{R_I} (R_I - \epsilon)$ and ending with the Hamiltonian $H_R (0) $, for the central point of the N-cube and to the Hamiltonian for fixed static coordinates for all the other points in the cube. $R_I - \epsilon$ is the position of the the stretched horizon.} The overlap rules must be compatible with the idea that the new degrees of freedom that are coupled in at some point on the lattice, can be thought of as ``coming from the neighboring points as they come into the horizon". The count of degrees of freedom says that the number of copies of the inflationary horizon is just the ratio of entropies, but in a normal inflationary picture this number is just $e^{3N_e}$ . If we assume that the scale of inflationary energy density is the unification scale, then this gives about $85$ e-folds, and the number is proportional to the logarithm of the ratio between the inflation scale and the scale of the asymptotic c.c.. Lower inflation scales give fewer e-folds. This is a sharpened version of our previous remarks on {\it a priori} holographic bounds on the number of e-folds\cite{numberofefolds}. We will call the lattice with Planck scale spacing {\it the fine grained lattice} and the lattice of central points of tilted cubes of size $n$, the {\it coarse grained lattice}.

Let us pause to emphasize a {\it very} confusing point. We are thinking of the lattice of observers of the HEI model in two conceptually different ways. In the HEI model itself, following the general rules of HST, the fine grained lattice represents different complete descriptions of all of cosmological history, which are constrained to agree about shared information. In our new model, we will consider that each fine grained point in the tilted N-cube has a Hilbert space of dimension $e^{\pi N^2}$, but we view its degrees of freedom as $e^{3N_e}$ copies of the degrees of freedom at each point of the coarse grained lattice of the HEI model. We will use the coarse grained lattice as a guide to constructing local interactions between the different copies. All of this is done at {\it each} point of the fine grained lattice of our new model. In principle, we have to supply overlap rules for every pair of such points, within the fine grained tilted N-cube, in order to make a consistent HST model of the universe. We will not do that in this paper, but instead concentrate on the universe as seen from a coordinate system built around the central observer in the N-cube, with the assumption that it evolves to a particular geodesic observer in the dS space with c.c. $\Lambda$. Other points in the cube evolve to accelerated observers at fixed static coordinates in that dS space , with the points at the boundary identified with stretched horizon observers, which see rapid scrambling of all $N^2$ degrees of freedom . For these boundary points, the time dependent Hamiltonian evolves directly from the sum of identical static Hamiltonians $H_{R_I}$ for the $e^{3N_e}$ copies, to the single static Hamiltonian $H_R$ of the final dS space. The interactions at the boundary are not local on the coarse grained lattice, and the time for this transition is of order $N$. We will describe the Hamiltonian for the center of the N-cube below, but will not complete the model by supplying a full set of Hamiltonians and overlap rules for all the points on the fine grained lattice on the N-cube. For large times, they should approach the rules for dS space of radius $R$, which we adumbrated above.

This model addresses from the outset the issue of ``trans-Planckian degrees of freedom" that has been discussed in the inflation literature. In the conventional picture, where our universe begins as a patch somewhat bigger than the inflationary horizon size, completely described by QFT, most of the localized degrees of freedom we see in the sky begin their lives as field theory fluctuations with wavelength much shorter than the Planck scale. Inflation theorists admit this but invoke the adiabatic theorem to argue that the wave function for these fluctuations is the Bunch-Davies vacuum. This amounts to a massive fine tuning of the initial conditions, which invalidates inflation's claim to explain the initial conditions of the universe that are required to explain observations. For a large quantum system, with a typical Hamiltonian, the adiabatic theorem is useful only for very special states near the adiabatic ground state. Typical level spacings in the generic part of the spectrum are too small for adiabatic arguments to work. The conventional inflationary argument for why initially trans-Planckian modes are in the Bunch Davies state, independently of their initial state, is, in our opinion, {\it wrong}.

In our models, homogeneity, isotropy and flatness are obtained for generic initial states, without inflation. The degrees of freedom are explicitly identified and their dynamics up through the inflationary era is completely modeled. As we will see, this leads approximately, modulo a crucial assumption, to the Bunch-Davies predictions for the spectrum of fluctuations, if we supply the appropriate time dependent Hamiltonian for the transition between the era when only $n^2$ degrees of freedom are coupled together and final state of the system.

Our model of the universe is a perfectly consistent HST up to a cosmological time of order the inflationary Hubble scale $n$.
After this point, we want the observers inside a tilted cube representing the Hubble scale of the c.c., $N$, to evolve to observers in the static patch of dS space. Outside that cube, we want, the system to evolve to the DBHF\footnote{We will discuss the effect of disjoint dS patches later. Their only utility in the model is to give an anthropic framework for choosing the c.c., and to solve the somewhat academic Boltzmann Brain problem.}. We know that the dS black hole in the $p=\rho$ FRW is a completely consistent solution of the rules of HST. However we now want to have an intermediate period in which for some period we have $(N/n)^2$ independent horizon volumes of the HEI model, which then interact locally to give the universe we observe and only approach the dS black hole solution in the asymptotic future. It seems likely that the dynamics on the boundary of the tilted cube of size $N$ will have to approach the final dS state more rapidly, in order to remain consistent with the DBHF outside, with which it has overlaps at all times. We have to find interior dynamics, which is consistent with a much slower approach to the dS vacuum state in the interior. This would amount to a derivation of approximately local dynamics from the HST formalism, and we have not solved that problem completely. We will outline a sequence of steps, concentrating only on the Hamiltonian at the central pointed of the N-cube.

We take the initial state of the system at the end of inflation, to be a tensor product of states in the Hilbert spaces of the individual observers at the centers of disjoint tilted cubes in the HEI. This is certainly consistent with the known dynamics of the system through the inflationary era. The individual factors are the same state initially, because of the consistency conditions. During the inflationary era, that state is acted on by the static Hamiltonian $H_{R_I} (R_I - \epsilon)$ in a way which thermalizes the system in a time of order the e-folding time. The time averaged density matrix is proportional to the unit matrix for each factor, and the tensor product density matrix is the maximally uncertain one on the full space-time, which is the ensemble corresponding to the dS vacuum of $H_R$. However, the microscopic pure state of the system is different at different times. Our fantasy model now, effectively\footnote{By effectively we mean that in the HST formalism, the Hamiltonian refers only to observables on a given time-like trajectory. Our decoupled systems correspond to different time-like trajectories. The essence of bulk locality is that information that is coupled into a given observer's Hamiltonian, can be viewed as coming from signals sent by other observers. We are not sure that we've learned how to do this in the HST formalism. What follows is our current best attempt.}, couples together these decoupled systems, producing the end of inflation. If that process occurs at slightly different times in different horizon volumes, then there will be local fluctuations in observables from horizon volume to horizon volume, which, by ergodicity, mirror the thermal fluctuations in the ensemble. Since the system has $e^{\pi n^2} $ states, with large $n$, we can expect the size of these fluctuations to have a Gaussian distribution, with root mean square $\frac{1}{n}$. This is a consequence of the central limit theorem. This is certainly true for the local energy density, which we take to simply be the Hamiltonian in each individual subsystem.

We'd now like to argue that we can make a model in which the two point function of the fluctuations is approximately invariant under the de Sitter group. Our basic strategy is to take $e^{3N_e} = (N/n)^2 $ copies of the degrees of freedom of one observer in the HEI model, and write a single observer Hamiltonian in this larger space, which is approximately local and dS invariant in the limit that $N_e$ is large. The locality of the Hamiltonian will be motivated by the space-time picture of the HEI model. To show that the model is consistent with the rules of HST, we must provide a set of overlap rules, between different observers, and we have not yet done this. We label the $(N/n)^2$ copies of the degrees of freedom of a single observer in the HEI model by the lattice points at the centers of each tilted cube of size $n$ in the space-time picture of that model. We call this lattice of centers, the {\it coarse grained lattice}. We want to construct a set of operators which, in the $\frac{N}{n} \rightarrow\infty$ limit, give rise to the dS group for the inflationary dS space. We have an $SU(2)$ subgroup of $SO(1,4)$, and we want to argue that, at least when $N_e$ is large, that this will extend to the full dS group.

We do this by identifying the points in the coarse grained lattice with points on a 3 sphere, according to the following construction. The metric of a three sphere is $$ds^2 = d\theta^2 + \sin^4 (\theta ) d\Omega_2^2 .$$ We will model a two sphere of radius $r$ in units of the inflationary Hubble radius as a fuzzy sphere.
That is, we assign a $p(p+1)$ matrix (and its conjugate) of spinor variables. For fixed matrix element, these variables satisfy the algebra describing a single inflationary Hubble volume, and different matrix elements anti-commute. The relationship between $p$ and $r$ is $\frac{r}{n L_P} = p$. We have a collection of such spheres with $p$ between $1$ and a large integer $P = e^{N_e}$. We arrange them along the $\theta$ interval $[0, \frac{\pi}{2} ]$, according to
the rule $$\sin^2 (\theta_p ) = p \sin^2 (\theta_1),$$ with $$\theta_P = \frac{\pi}{2} .$$ The other hemisphere, covering $[\frac{\pi}{2}, \pi ]$, is described by a second copy of the same set of degrees of freedom. We view it as the thermofield double Hilbert space.

To localize points on the fuzzy spheres, we note the identity
$$\delta (\Omega , \Omega_0) \delta_{ab} = \sum _L\ \psi^*_{a L} (\Omega_0) \psi_{b L} (\Omega ) .$$ The label $L$ runs over the spinor spherical harmonics. If $L$ is restricted to be $\leq p(p+1)$ by putting a bound on the Dirac eigenvalue, then we get a fuzzily localized function $\delta_p (\Omega , \Omega_0 ) $. This set of functions is over-complete in the space spanned by the cutoff Dirac eigen-sections. However, if we restrict attention to a discrete set of points lying at the centers of the tiles of an icosahedral tiling of the sphere, then it is clear that, as we increase the number of tiles, we go from an incomplete to an over-complete basis of localized functions. For large $p$ we make no mistake by taking the minimal over-complete basis. Note also that as $P \rightarrow \infty$, $\sin^2 (\theta_1 ) \rightarrow 0$, so we get a set of variables associated with a dense set of points on the 3-sphere.

We can define an action of $SO(4) = SU_L (2) \times SU_R (2) $ on our system by noting that a $p \times p+1$ matrix has a natural $SU(2) \times SU(2)$ action
separately on its rows and columns, such that it transforms as the cutoff spinor bundle on the two sphere under the diagonal $SU(2)$\footnote{There is no analog of this in higher dimensions. We believe that this observation is connected in an as yet mysterious way to the fact that supergravity has no de Sitter solutions in more than $4$ dimensions, whereas in $4$ dimensions we can easily build dS solutions using the superpotential of chiral multiplets.}.
We can now build a lattice field theory out of our spinor variables on the discrete finite approximation to the 3-sphere that we have constructed. As usual, when $P \rightarrow\infty$ and the points become dense on the sphere, we can get a continuum limit by insisting that the system be invariant under the conformal group of the sphere, $SO (1,4)$, which is the de Sitter group. We define the Hamiltonian by insisting that it be the $SU(2)$ invariant boost generator of $SO(1,4)$. $SU(2)$ is a symmetry even for finite lattice spacing, while the rest of the dS group is an IR symmetry. Note that since the parity of $p$ changes between adjacent spherical shells, products of inter shell nearest neighbor pixel operators will contain the four vector representation of $SO(4)$ and provide candidate lattice Hamiltonians which approach the dS generator.

We do not know how many different fixed points can be accessed from our particular set of lattice variables. It may be that the obvious precursors of the $SO(4)$ action (transformations that map discrete points in our tiling into each other, accompanied by the corresponding rotations on the matrix rows and columns)
forces us to be in the basin of attraction of a particular fixed point\footnote{A possible candidate would be a Chern-Simons theory coupled to 3 dimensional massless fermions.}. At the moment, our considerations are too rough to see any phenomenological difference between different fixed points, if they exist.

Let us now reiterate our goal and summarize our progress towards achieving it. We have constructed two model cosmologies called HEI and the dS black hole in the $p = \rho$ universe. They both live on a cubic lattice with Planck spacing. In HEI we satisfy the rules of the DBHF cosmology up to a time $n$ and then stop the growth of the Hilbert space and evolve forever with a rescaled Hamiltonian. The integer $n$ is chosen large enough that the scaling rules of the $n \rightarrow \infty$ limit are approximately true. The resulting state, at every point of the lattice rapidly cycles through the Hilbert space, whose dimension is $e^{\pi n^2}$. The time averaged density matrix is maximally uncertain. Thermal fluctuations, which, by ergodicity, mirror the fluctuations in time, are of order $1/n$ and approximately Gaussian. This is true for a large set of operators, including the Hamiltonian. The overlap rules of the HEI model imply empty overlap at all times for points that are separated by more than $n$ lattice spacings. Thus, we can consider as independent, the degrees of freedom associated with the centers of disjoint tilted cubes, which are the locus of all points $\leq n$ lattice steps from a given point. This new lattice of centers is called the coarse grained lattice.

We have then shown how to organize $e^{3 N_e}$ of these coarse grained points into a system, which becomes $SO(1,4)$ invariant, up to exponential corrections, as $N_e \rightarrow \infty$. We consider the point at $\theta_1$ to be identified with the observer trajectory corresponding to a specific point on the original fine grained lattice. Its Hilbert space has entropy $e^{3N_e} \pi n^2 \equiv \pi N^2 $. Its time dependent Hamiltonian is the sum of the Hamiltonians for the original fine grained center points, up until time of order $n$, and then makes a transition on the same time scale to the de Sitter boost Hamiltonian.
$$H_I (t) = \sum_{\Omega_{p,k} } H_{p,k} (t) \rightarrow J_{04} . $$ $\Omega_{p,k}$ label points on the hexagonal tilings of all of the spherical shells in our construction of the 3-sphere. The label on this Hamiltonian indicates that it describes dynamics up through the inflationary era of the universe. The transition to $J_{04}$ takes place at different times at different $p,k$, so our initial state is inhomogeneous, with small Gaussian fluctuations from point to point. In order to be consistent with causality, the Hamiltonian must gradually couple the degrees of freedom at $\theta_1$ to those on larger shells, and this requires the space dependence of the transition. The fluctuations arise from the ergodic dynamics of the individual Hamiltonians $H_{pk}$. For large $N_e$ we can now invoke approximate $SO(1,4)$ invariance to argue that the two point function of the fluctuations will have the usual scale invariant form. We have not yet understood the nature of the corrections to dS invariance, though we can certainly see that fluctuations will be suppressed in both the extreme UV and the extreme IR. The corrections to Gaussianity are down by powers of $n$. We further assume that, as in QFT in dS space, we can write $H_I (t)$ as a difference of two identical terms each of which involves only degrees of freedom in one hemisphere. We view the hemisphere at $\theta \leq \frac{\pi}{2}$ as the physical Hilbert space and the other hemisphere as it's thermofield double.

Now let us turn to the dS-DBHF model. This uses one tilted cube of $N$ Planck sized steps, inside of which we have the rules of HEI with Hubble radius $N$.
Outside this cube we have the rules of the DBHF model, with its asymptotically infinite Hilbert space. There are no overlaps between points on the interior and exterior of the cube\footnote{These points are thus causally disconnected forever.} The overlaps between exterior points and the points on the boundary of the cube grow according to the DBHF rules until time $N$, and then the overlap is the entire finite dimensional boundary Hilbert space. This indicates that the boundary of the cube is a stable marginally trapped surface. For large $N$, the ``carpenter's ruler" nature of the overlap rules shows that this surface has the geometry of a two sphere of radius $N$.

We now want to replace the dynamics at the center point of this cube by something that resembles, for a time, the Hamiltonian $H_I (t)$, restricted to the physical hemisphere. This introduces an approximate locality and dS symmetry into its dynamics. {\it This cannot last forever.} The boundary points of the N-cube must have the dynamics of the maximally accelerated static observer in dS space, in order to satisfy the overlap conditions with the exterior DBHF.
We conjecture that this is only possible, consistent with the stringent rules of HST for the entire fine grained lattice, if the interior points of the cube have asymptotic dynamics corresponding to some time-like trajectory in the dS space with Hubble radius $N$. For the center point, which we'd like to identify with our own trajectory through space-time, this should be, to a good approximation, a geodesic. The Hamiltonian $H_I (t)$ must therefore evolve, over a time scale shorter than $N$ into the geodesic Hamiltonian.
$$H_G = P_0 + \frac{1}{N} V ,$$ where $V$ is a fast scrambler Hamiltonian\cite{fastscramble} with $|| V || \sim 1$. $P_0$ is the bilinear Hamiltonian which appears in the SUSY commutation relations of asymptotic particle states. It decouples the particles in the $O (N^{1/2})$ dS horizon volumes which are allowed by the theory of stable dS space. On time scales $\leq N$, this Hamiltonian gives rise to a transition operator between incoming and outgoing particle states, while on longer time scales it thermalizes the entire system, sweeping over its full Hilbert space. This maximally uncertain density matrix is thermal at the dS temperature for the Hamiltonian $P_0$ because of the relation between degeneracies and eigenvalues of $P_0$\cite{dS}. The full story of our inflationary cosmology is thus a time dependent Hamiltonian that makes transitions as follows

$$H(t) = H_{DBHF} \rightarrow\Sigma H_{dSn} \rightarrow J_{04} \rightarrow H_{cosmo} \rightarrow H_G ,$$

The timescale for the first two transitions is of order the inflationary Hubble scale $n$, whereas the last one takes a time of order $N$. The two middle Hamiltonians describe the generation of fluctuations during the inflationary era. The next to last transition is the reheating period and $H_{cosmo}$ describes cosmic history during the radiation and matter dominated eras. We do not have an HST description of this era. Furthermore, we have described the inflationary era by a Hamiltonian corresponding to a single observer, but have not yet found a complete model satisfying all of the overlap rules of HST.

It's interesting to compare and contrast our model with the conventional picture of inflation.

\begin{enumerate}

\item The origin of local fluctuations, as in conventional inflationary models, is a time mismatch between different parts of the system.
Just before the transition to the Hamiltonian $J_{04}$, the initial density matrix of the system is the product of maximally uncertain density matrices in the individual inflationary horizon volumes. This is the same as the maximally uncertain density matrix in the full Hilbert space, so in a coarse grained way, the system is always close to the final dS vacuum. However, the initial pure state is a tensor product of states chosen from the individual horizon volume Hilbert spaces, and this is certainly {\it not} a typical state in the full Hilbert space. It is also different from place to place if there is a mismatch of the times at which the transition between the two Hamiltonians is made.
Note that the localization of the fluctuations is a consequence of the intermediate era in which the model is described by the rules of the HEI model, which can be broken into tilted cubes, at the center of each of which we have an independent set of degrees of freedom. The fluctuations have a size determined by the dS temperature, and are small and Gaussian because this temperature is small in Planck units (equivalently, the entropy $\pi n^2$ is large).
On the other hand, the quantum degrees of freedom are not a scalar field, and there is nothing quantum mechanical about the fluctuations. They are purely thermal in origin, coming from the ergodic dynamics of $H_{pq}$ and the time mismatch of the transition at different points. We note in passing that no observable consequence of inflation can ever detect the quantum nature of the fluctuations. Indeed, an important part of the usual story of inflation is the decoherence of different sectors of the wave function, describing macroscopically different patterns of fluctuation, because the approximately classical dynamics of these variables is coupled to the quantum excitations of all the particles in the universe. With good accuracy, the conventional picture of inflation is model by a stochastic classical field theory, coupled to gravity.

\item It is conventional to cut off inflationary fluctuations at a coordinate length scale equal to the Hubble scale. This is a time-dependent cutoff on physical scales and allows ``trans-Planckian" modes to appear at late times. Our lattice field theory has a physical cutoff at the inflationary Hubble scale, on a sphere whose size already takes into account the inflationary expansion. Thus, viewed from a conventional perspective, it contains such ``trans-Planckian" modes. However, our perspective is not the conventional one. We have a complete microscopic theory of quantum gravity and we have described a time dependent Hamiltonian for the fundamental variables, which produces a situation very similar to the final state of inflation. There is nothing mysterious at the Planck scale in our model, and we have made no specious adiabatic argument to get around the mystery. If we succeed in completing our program, by really proving approximate dS invariance, and providing a set of Hamiltonians for all the other observers on the fine grained lattice, which satisfies the overlap constraints of HST, we will have constructed a model which predicts localized fluctuations which are small and approximately Gaussian and scale invariant. We are not yet in a position to make intelligent comments about the possible red tilt of the spectrum.

\item We have not yet provided a theory of reheating. Indeed we have not yet built a model for the central point in the N-cube, which is consistent with the basic requirement that the Hamiltonian approach the geodesic Hamiltonian $H_G$ for the large radius dS space. The, necessarily time dependent, evolution between $J_{04}$ and $H_G$, is the analog in our formalism of the period of reheating. This will tell us what the reheating temperature is, whether there is a period of matter dominated expansion driven by homogeneous scalar field oscillations, whether a baryon asymmetry is generated during this era, {\it etc.}. The excitations of $H_G$ are the particles we see in the present universe and the reheating process tells us many things about our expectation for their distribution. At the present stage of our knowledge we cannot say much about this period. In conventional inflation models the details of the reheating period depend in a crucial way on the Lagrangian for the inflaton field (a beast which has not yet put in an appearance in our discussion). We suspect that the overlap constraints on HST models, combined with the fact that these models have a fixed set of degrees of freedom (whereas inflation models might have multiple fields, or non-standard Lagrangians ), will make the predictions for reheating more definite, but we cannot claim with any plausibility that they will be unique, or that we know what they are. One conventional inflationary prediction is quite independent of the details of reheating, as long as the transition between inflation and more conventional expansion is relatively abrupt. This is the one parameter prediction of the spectrum of gravitational waves. It depends only on the Bunch-Davies two point function for gravitons, and on the applicability of the sudden approximation. This is due to the fact that gravitons are so weakly coupled. Our model definitely has gravitons, and we have given the beginnings of an argument for approximate dS invariance, so it is plausible that it will make approximately the same predictions for the gravitational wave power spectrum.

\item Jacobson's arguments indicate that there should be an effective classical field theory, including gravity, which describes the space-time of our model. We have already seen that the HEI model, and the dS black hole in the DBHF require a scalar field to model space-time variations of the cosmological constant, so it is extremely plausible that the required model is a slow roll inflation model, with a stochastic but small inhomogeneity in its initial conditions at the end of inflation. The stochasticity arises from the mismatch of local transition times in the quantum model, between the decoupled Hamiltonians of individual horizon volumes, and $J_{04}$. To be pedantic, we are actually defining a whole class of models, each with a different distribution of inhomogeneities, and making statistical predictions about the consequences of a generic model in the class. The models are defined by the spatial distribution of times at which the local single horizon Hamiltonian evolves into the Hamiltonian $J_{04}$. A marked difference between conventional slow roll inflation models and the HST model of inflation is that here we must compute the fluctuation spectrum from the underlying quantum model, rather than simply apply perturbation theory to a Lagrangian quantum field theory of the inflaton. The fluctuations are thermal rather than quantum\footnote{Of course, the fluctuations in the Bunch-Davies state are a mixture of thermal and quantum fluctuations.}. We cannot simply assert that the fact that the thermodynamic effective field theory is the same as slow roll inflation, implies that the fluctuation spectrum is the same. In fact, in the HST model {\it the inflaton is not a quantum field}. It is most assuredly not `` a field corresponding to a particle in the underlying string theory". Particle excitations have meaning in the HST formalism only for very large causal diamonds and for times when the Hamiltonian is approximately $P_0 (r)$, where $r$ is the radius of the diamond. This Hamiltonian decouples the particle excitations from the rest of the degrees of freedom in the diamond. This is certainly not the case during the time of inflation, when, according to our model, the covariant entropy bound is close to being saturated. In such situations, Jacobson's arguments indicate the existence of a classical effective field theory, but quantized field theory is not a good approximation to the real dynamics.

\item Conventionally, one says that inflation explains the homogeneity, isotropy and flatness of the universe, without fine tuning initial conditions. This argument is incorrect. Its justification in the literature is the adiabatic theorem. However in generic systems the adiabatic theorem works only if the initial state is very near the ground state, and such a choice is a fine tuning of initial conditions. In HST models homogeneity, isotropy and flatness all follow for generic states from the rules of the formalism. We cannot satisfy the consistency conditions without imposing homogeneity and the overlap rules become rotation invariant, in the same limit in which space-time emerges from quantum mechanics. The role of inflation is instead to VIOLATE homogeneity. The inflationary era of our model is the reason that the model contains localized fluctuations of the asymptotic dS ground state. We can parameterize this in terms of the integers $n$ and $N$, since we have a different model for each choice of these integers satisfying $1 \ll n \ll N$. The magnitude of the fluctuations is determined by $n$ and the cutoffs on the scale invariant spectrum by $N/n$.

\end{enumerate}

We have tried to make clear the logic of how the 3 well defined models of HST, which we have described above, are fit together in our heuristic model of the universe. In the interest of what we hope is further clarification, we append the following schematic diagram of the relations between the models:

\begin{figure}[h]
\begin{center}
\includegraphics[width=7.5cm]{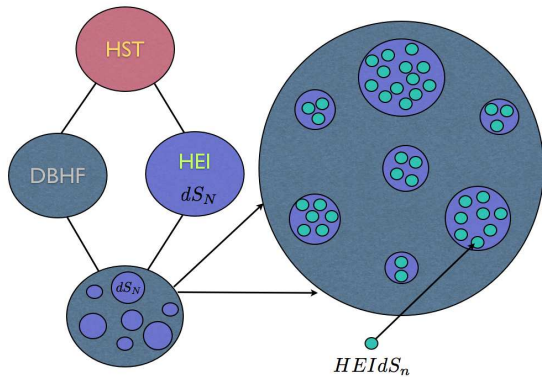}

\end{center}
\caption{Schematic relation between the different cosmological models. The left chart shows 3 models satisfying all the rules of HST. The infinite DBHF and HEI models, and the black hole in the DBHF, with a dS interior. The right chart is our heuristic idea of how these models are put together to make a more realistic cosmology in which an early era of inflation generates localized excitations in a dS universe, which is embedded in a DBHF-dS multiverse.}
\end{figure}

\subsection{Why the universe had low entropy}

If a consistent HST model of the type we have just discussed can be constructed, then it predicts that as we approach the asymptotic dS space, we
have a localized excitation of the vacuum. Our discussion of dS space indicates that this is a lower entropy state than the dS vacuum.
{\it The more entropy we have
in localized degrees of freedom, the smaller the entropy of the
total state of the system.} In other words, from the point of view
of the final high entropy state of the system, the most probable way is
to make a transition between the DBHF and dS space with c.c.
$\Lambda$ which has no localized excitations at all at intermediate times. We have however constructed a model which does have such excitations.
Roughly speaking, such a model is characterized by two integers $n$ and $N$ \footnote{Strictly speaking, every distinct pattern of fluctuations is, in our language, a different model, but we have agreed to treat this ambiguity statistically, and are working
with a typical model.}, which determine the inflationary and asymptotic values of the c.c. There is apparently nothing preventing us from constructing
a consistent model with any two integers satisfying $ 1 \ll n \ll N$. If we go back to our picture of a collection of dS black holes in the DBHF, we can now think of each black hole in that collection being given the additional label $n$.

It is obvious now that, within such a
model, anthropic selection criteria are important in determining the
state of the universe we see. However, many of the conventional
arguments may be turned on their heads. For example, in this
context, Weinberg's\cite{wein} constraint on galaxy formation, $$
\Lambda < Q^3 \rho_0 ,$$ may have an interpretation very different
than the conventional one.

According to the hypothesis of Cosmological Supersymmetry Breaking,
which follows inevitably from the holographic formalism, the c.c. $\Lambda$
fixes the scale of SUSY breaking, via
$$m_{3/2} = K \Lambda^{1/4}.$$ Thus, all of low energy particle
physics depends in an interesting way on the relation between $\Lambda$
and the confinement scales of QCD and the hidden sector gauge group.
The parameter $K$ scales like the square of the Planck scale over
the unification scale. Dark matter is likely to be a hidden sector
baryon, with a density determined by asymmetries generated at the
unification scale. The low energy effective field theory of a stable
dS space may be fairly unique\cite{pyramid}\footnote{The $\Lambda = 0$ model is a super-Poincare invariant model of gravity with a discrete R symmetry and no moduli.
There are no known examples among putative string theory solutions. Furthermore, when one imposes phenomenological constraints, such as the standard model gauge group, coupling unification, and gaugino masses above current experimental bounds, the list of possible low energy effective theories becomes very small, and is reduced to the class of Pyramid Schemes.}. Thus, it is entirely plausible that anthropic considerations like the existence of nuclei and atoms completely fix $\Lambda$ and $\rho_0$ to be near their real world values. If this is the case, then Weinberg's galaxy formation bound becomes a lower bound on $Q$, $$ Q > (\frac{\Lambda}{\rho_0})^{1/3} .$$ Purely statistical considerations imply that the most likely model universe interpolating between the DBHF and a stable dS space with a given value of $\Lambda$ is the dS black hole in the DBHF fluid. A model with an intermediate scale of inflation has a less likely state during the inflationary era\footnote{The logic here is a bit convoluted. Each of our model universes corresponds to a different time dependent Hamiltonian evolution operator. For each choice of evolution operator, we make a generic choice of initial state consistent with the rules of HST. When we make a statement like "the state with this choice of evolution operator is less likely than the state with that choice", we are comparing two different models of the universe whose final state is the same maximally uncertain density matrix. One model forces the universe to evolve through a more special class of intermediate states than the other.}. Thus, the statistical preference for a model with $Q=0$, which really means a model with no intermediate inflationary era, is countered by the anthropic selection of universes in which galaxies can form. If we look for the minimal value of $Q$ consistent with a galaxy, we overestimate the true value of $Q$ by a factor of $10^{2/3}$. More refined anthropic arguments, like those in \cite{boussoetal} might reduce this mild discrepancy.

One often asks why anthropic considerations do not predict that only a single galaxy is
formed, in an otherwise empty universe. Our model does not appear to have a mechanism for such a process.
It builds the observed universe out of components whose degrees of freedom are all in equilibrium. The only universe we can make with a single localized
object in it, gives us a black hole in dS space. We could wait for a black hole recurrence time, for the black hole to spontaneously fluctuate into a galaxy of equal mass, except that the black
hole evaporates on a much shorter time scale. So the only single galaxy universes that a model of this type predicts are those where the galaxy appears
once every dS recurrence time.

The actual model we are proposing will have a different Hamiltonian and overlap rules than such a single galaxy universe, and will produce structure in a more or less normal CDM manner. By virtue of its dynamics, it produces small localized inhomogeneous fluctuations in a roughly homogeneous universe, which can only form galaxies
in the conventional manner, and is guaranteed to produce many of them, if it produces any at all.

\subsection{Boltzmann Without Brains}

In recent years the community of inflationary cosmologists has concerned itself with the so-called problem of Boltzmann's brains. According to this argument a model in which the universe comes into thermal equilibrium at a fixed temperature is inconsistent. If one is enamored of anthropic arguments, then one is tempted to ``count intelligent observers", and argue that the universe we observe should be that seen by the ``typical observer in the total ensemble of universes in the multiverse". We see no particular reason to adopt this hypothesis. For example, if it were really true that theory predicted a multitude of universes with different color gauge groups, and that the typical group was $SU(5)$, we would simply say that the fact that we have $G = SU(3)$ could just be an accident. Even if the probability distribution had the form $e^{ - c (N - 5)^2},$ with fairly large $c$, we could simply swallow our disappointment that we were not a part of the in-crowd and proceed to do physics by looking to see whether we were typical among all those universes, which had $N = 3$.

The only truly compelling version of the Boltzmann brain argument is the one originally invented by Boltzmann's assistant Schutz, to shoot down Boltzmann's idea that the reason we see the second law of thermodynamics in operation is that the universe began as a low entropy Poincare recurrence in a finite system with a time independent Hamiltonian. If our explanation is such a fluctuation, then there are much less drastic fluctuations in which an intelligent observer is nucleated with all of its memories. Such an observer is destroyed in a relatively short time, but its instantaneous experience is identical to our own. Arguing statistically, we would (each individually) argue that in such a model of the history of the universe, we are most probably such a fluctuated observer, but our continued existence shoots down this hypothesis repeatedly. Versions of this argument have resurfaced recently in the context of a theory of the universe in
which it began as a fluctuation in an eternal dS space with finite entropy\cite{dks}.

We think it is a mistake to apply this logic to a system which does not postulate Boltzmann's explanation of the second law of thermodynamics, particularly one like HST which has a time dependent Hamiltonian. The purpose of theoretical physics is to explain data, and the data we can hope will be available even to the most fantastical future race of intelligent computers is limited by the characteristics of the universe. The problem with using the Boltzmann brain argument is two-fold. First, in a system with time dependent Hamiltonian we can modify the future evolution of the Hamiltonian to make the BB problem go away, in an infinite number of ways which have no effect on the model's predictions for the universe we observe, or could possibly observe in the future. We will present an explicit mechanism for this in the context of our HST model in a minute.

The claim that our modifications will not effect observations is based on the time scale that it takes for a BB to be nucleated. In thermal equilibrium at the dS temperature, the probability of nucleating a brain small compared to the Hubble radius is of order
$$e^{ - \frac{2\pi 10^{61}\ m}{M_P} - \frac{\pi m^2}{M_P^2} } .$$ The first term is the thermal probability of nucleation of a system of mass $m$ and the second the probability that that system is not a black hole. We have deliberately left the units off this rate per unit volume per unit time, since the answer is essentially the same number measured in Planck units as it is in ages of the universe, even if we imagine an almost certainly impossible intelligent observer with mass equal to the Planck mass.

The second term in the exponent is negligible compared to the first for many purposes, but it makes an important point. If we imagine a future civilization waiting to observe the nucleation of a BB\footnote{It's impossible to imagine {\it an intelligent} future civilization doing this.} it will see a huge number of black holes nucleated before the BB, most of them microscopic and giving off deadly bursts of Hawking radiation. Thus, the probability that one could actually see a BB is incredibly small.

In fact, there is, in a finite dimensional model of dS space, a fundamental barrier to observing {\it any } sufficiently improbable Poincare recurrence. In order to interpret the mathematical formulae of quantum mechanics in terms of experiments, we have to postulate the existence of large systems, which approximately obey the rules of quantum field theory. A quantum measurement consists of the entanglement of a microscopic quantum state with a macroscopic variable. Macroscopic variables are coarse grained averages of quantum variables in large systems. They have very small intrinsic quantum fluctuations, which are inversely proportional to the number of degrees of freedom in the system. More importantly they represent a huge ensemble of quantum micro-states, all of which have the same classical history, and among which the system makes transitions on a time scale rapid compared to the classical motion. This leads to a decoherence of quantum phases of the macroscopic variables state which is exponentially small in the number of degrees of freedom.

In the quantum theory of gravity, there are two kinds of large quantum system with macro-observables: black holes, and systems well described by QFT. The latter can have many different macro-observables. They can exhibit complex classical behavior and store large amounts of information in a robust way. Black holes, as a consequence of no-hair theorems, have very few macro-observables and do not make good measuring devices. In the HST theory of stable dS space, the overwhelming majority of states correspond to empty dS space. They are localized on the cosmological horizon and form a featureless equilibrium system, with no classical observables besides the thermodynamic properties of dS space. Localized systems are low entropy excitations of the theory, and can make good measuring devices. But the number of quantum states that can be made without forming a black hole of size comparable to the horizon is only $e^{N^{3/2}}$. This is a wild overestimate of the information storing capacity of any localized device in dS space, which means that {\it in principle} the theory does not allow for measurement of the quantum state of the horizon. Conversely, there will be many choices for the Hamiltonian of dS space, which make the same predictions for all possible measurements, but different predictions for the quantum state of the horizon.

Over a time of order no longer than $e^{N^{3/2}}$ (where again, for $N \sim 10^{61}$, the units are irrelevant), any would be measuring device in dS space will undergo uncontrollable quantum fluctuations, so there is no operational meaning to observing events for which the probability is smaller than $e^{- N^{3/2}}$.
Thus, Boltzmann recurrences of objects of mass $> 10^{30} M_P = 10^{22} kg $ which is about $1\%$ of the mass of the earth, are in principle unobservable.

The problem of Boltzmann's brain thus involves ridiculously long time scales, which have no place in a serious discussion of science. To illustrate how HST could resolve the problem, recall that our anthropic explanation of the c.c. requires us to envision a large collection of black holes in the DBHF, each with a dS interior and a different c.c. . According to Einstein's equations the evolution of this system depends on the initial positions and velocities of the black holes. Any set of positions and velocities which leads to a black hole of our observed c.c., which remains in isolation for time scales much longer than its Hubble radius, but then collides with another black hole of comparable or larger radius, will solve the BB problem. The collision will lead to a black hole with larger radius. Its interior will either be singular, or resemble a dS space with c.c. much smaller than our own. In the first case there are no Boltzmann brains because the singularity is reached in times exponentially shorter than the nucleation time. In the second, given the relation between the c.c. and SUSY breaking, the scales of particle physics will be drastically different than they are in our world, as will their ratios. Nuclei and atoms will not exist and there are again no Boltzmann brains. Notice that there are an infinite number of models of this type, which resolve the BB problem, without effecting any experiment that we can ever do. If a super-civilization could last for times much longer than the current age of the universe, we might have to modify our initial conditions to take into account their improved observations. However, since BB nucleation probabilities are so much smaller than any (inverse) cosmological time scale, we can easily do this without encountering a problem.

\section{Conclusions}

We've outlined a theory of holographic space-time, which generalizes
both field theory and string theory and sheds light on many of their
defects. It provides a frame-work in which SUSY inevitably emerges
in Poincare invariant limits. Moduli are also revealed to be
emergent properties of Poincare invariant or AdS limits, in which we've allowed certain integers characterizing the
fuzzy internal geometry of HST go to infinity. They are not relevant for cosmology.

Large radius dS space is a finite dimensional quantum system with no
moduli. It is described by a one dimensional discretuum of Hamiltonians,
corresponding to fixed values of the static spatial coordinate.
Among these, the geodesic observer has the largest ratio between the free particle Hamiltonian
and the fast scrambling Hamiltonian, which thermalizes all of the degrees of freedom of the system.
On times scales shorter than the Hubble scale, we can define an approximate scattering amplitude for the particles.
States in which the particles are unentangled in the past, get mapped approximately into different unentangled
states in the future. On time scales long compared to the Hubble time, the system is completely thermalized and explores
its entire Hilbert space. From the point of view of the free particle Hamiltonian $P_0$, this corresponds to a finite temperature $(2\pi R)^{-1}$, as
a consequence of the relation between the degeneracy of the $P_0$ eigenspaces, and the eigenvalues themselves.
The dS vacuum is the
infinite temperature density matrix of this system and localized
excitations of it are lower entropy excitations, with the total
entropy going down as the localized entropy increases.

We also reviewed and improved a model of early universe cosmology. We
now have quantum mechanical models describing an infinite dense
black hole fluid with $p = \rho$, and a DBHF that evolves to an inflationary universe
with fixed c.c., and an infinite number of decoupled horizon volumes. Finally we have a model in which the DBHF develops a
spherical defect, which looks like a black hole from the outside and dS space on the inside.
We've also described a possible variation
of this last model, for which there is as yet no complete microscopic
definition, In this hypothetical model, there is an intermediate stage with $N_
e \sim \frac{1}{3} {\rm ln} \frac{\Lambda_I}{\Lambda} $, e-folds of
inflation, which intervenes between the DBHF and the asymptotic dS
space. We presented an argument that this model could enable us
to come to an understanding of why the localized entropy of our own
universe was low. The argument combines the observation that, in
dS space, increasing the localized entropy decreases the total
entropy, with Weinberg's anthropic relation between the c.c., the
primordial density fluctuations, $Q$, and the initial dark matter
density. If we assume the c.c. and the dark matter density are
fixed by other considerations, this relation is a lower bound on
$Q$. Our remark about localized and total entropy implies that the
{\it a priori} probability distribution for all models which
interpolate between the DBHF and dS space, favors small $Q$. Weinberg's lower bound fixes $Q$ to be close
to its value in the real world.

Anthropic arguments have a somewhat different flavor in HST models than they do in conventional points of view. Each different coarse grained history of the universe is, in this framework a different (time dependent!) model Hamiltonian. Different initial states for the same Hamiltonian do not lead to substantially different histories because we never stray far from equilibrium for the global degrees of freedom. Our model of black holes embedded in the $p=\rho$ FRW, with different asymptotically dS black hole interiors, allows for different local evolutions, primarily characterized by the inflationary and asymptotic values of the c.c., which in HST are the two integers $n$ and $N$. There {\it may} also be a choice of super-Poincare invariant four dimensional zero c.c. limits of the theory of stable dS space. Assuming low energy physics like our own, an interesting picture emerges. It has been argued elsewhere that the only phenomenologically viable low energy models consistent with coupling constant unification are the Pyramid Schemes\cite{pyramid}. In the zero c.c. limit of these models, there is {\it no} asymptotically free gauge theory and SUSY is unbroken. The value of the c.c. fixes the R symmetry breaking operators, which determine the scale of SUSY breaking, and the confinement scales of both the Pyramid gauge group and QCD. It also effects the running of other standard model couplings. Finally, dark matter is a hidden sector baryon, whose properties depend on the Pyramid scale, and whose abundance is determined by GUT scale physics in the model. Thus, it is extremely likely that both the asymptotic c.c. and the dark matter density, are determined by a combination of mathematical properties of the model, and anthropic considerations like the existence of atoms. It's important to recognize that ALL of the properties of this effective field theory, including the GUT scale physics, are in principle determined by a single parameter, the asymptotic c.c. . The only anthropic parameter left is the inflationary c.c. and mathematical properties of the model constrain that to be much less than the asymptotic c.c. and much smaller than the Planck scale. If we consider different models of the universe with the same asymptotic c.c., from the point of view of the states in the asymptotic Hilbert space, then we maximize the total entropy by choosing {\it the smallest} inflationary c.c., leading to the smallest localized fluctuations, and the smallest deviation from the maximal entropy dS vacuum state. However, Weinberg's argument shows us that if the fluctuations are too small, {\it no galaxies are formed}, and localized entropy production, which would seem to be a {\it sine qua non} for any form of organized life, does not occur. The considerations of \cite{boussoetal} may enable us to strengthen this argument. Thus, galaxy formation is seen to be a biothropic lower bound on the inflationary c.c. . The most naive form of this bound is too small by a factor of order $5$ in $Q$ the measure of primordial fluctuations.

In the course of our investigation, we've uncovered a number of unusual properties of HST models. Inflation and dS space are very
different beasts. The number of e-folds of inflation in our hypothetical model of the real universe is fixed by the ratio between the inflationary and
asymptotic values of the c.c. . Effective classical field theory is always a valid coarse grained description of HST, but fields, like the graviton and inflaton fields
which describe situations where the covariant entropy bound is close to being saturated, are not to be quantized. Bulk quantum fields only arise in the description of the excitations
of asymptotically flat and AdS space-times (and perhaps others with similar asymptotic properties), and in the approximate description of large radius dS spaces.
We have also adumbrated a theory of accelerated observers and Unruh radiation, of which more will be said elsewhere\cite{holounruh} .

We are well aware that our theory of inflationary fluctuations is incomplete, and hope to return to a more satisfying treatment in the near future.

\section{Acknowledgements}

This work of WF was supported in part by the NSF (grant number PHY-0969020) and that of TB by the DOE.


\end{document}